\documentclass[iop]{emulateapj}
\usepackage{dsnr,epsfig,hyperref,ifthen,natbib}
\shorttitle{Oxygen Gradients in Interacting Galaxies}
\shortauthors{Rupke et al.}

\begin{document}

\slugcomment{Accepted to ApJ 25 Aug 2010}

\title{Gas-Phase Oxygen Gradients in Strongly Interacting Galaxies: I.
  Early-Stage Interactions}

\author{David S. N. Rupke\footnote{Present address: Department of
    Physics, Rhodes College, Memphis, TN 38112}, Lisa J. Kewley,}
\affil{Institute for Astronomy, University of Hawaii, 2680 Woodlawn
  Dr., Honolulu, HI 96822} \email{drupke@gmail.com}

\author{and L.-H. Chien} \affil{Space Telescope Science Institute,
  3700 San Martin Dr., Baltimore, MD 21218}

\begin{abstract}

  A consensus is emerging that interacting galaxies show depressed
  nuclear gas metallicities compared to isolated star-forming
  galaxies.  Simulations suggest that this nuclear underabundance is
  caused by interaction-induced inflow of metal-poor gas, and that
  this inflow concurrently flattens the radial metallicity gradients
  in strongly interacting galaxies.  We present metallicities of over
  300 \htwo\ regions in a sample of 16 spirals that are members of
  strongly interacting galaxy pairs with mass ratio near unity.  The
  deprojected radial gradients in these galaxies are about half of
  those in a control sample of isolated, late-type spirals.  Detailed
  comparison of the gradients with simulations show remarkable
  agreement in gradient distributions, the relationship between
  gradients and nuclear underabundances, and the shape of profile
  deviations from a straight line.  Taken together, this evidence
  conclusively demonstrates that strongly interacting galaxies at the
  present day undergo nuclear metal dilution due to gas inflow, as
  well as significant flattening of their gas-phase metallicity
  gradients, and that current simulations can robustly reproduce this
  behavior at a statistical level.

\end{abstract}



\section{INTRODUCTION} \label{sec:introduction}

The production and redistribution of heavy (non-hydrogen or -helium)
elements within galaxies is an important aspect of galaxy evolution.
The amount of heavy elements in the gas phase of galaxies has been a
topic of study over the decades since resolved galaxy spectroscopy
became commonplace.  In particular, emission line spectroscopy of
star-forming regions has been of great use in understanding the
distribution of chemical elements in nearby, normal galaxies.  For
more distant galaxies, the mass-metallicity relationship, an important
signpost of galaxy chemical evolution, has been used to constrain
chemical evolution across a broader range of galaxy types and,
increasingly, across cosmic history.

Intense study of the mass-metallicity (hereafter, \mz) relation in
recent years has yet to pin down its origin
\citep[e.g.,][]{tremonti04a,kwk06a,brooks07a,dalcanton07a,zahid10a}.
However, it is apparent from an increasing number of unique studies
that part of the scatter in the relation is due to a nuclear
underabundance in interacting galaxies, at both low
\citep{lee04a,ekta10a} and high masses
\citep{kewley06b,rvb08a,ellison08a,michel08a,peeples09a,solalonso10a}.
If one of the progenitor galaxies in an interacting system is a
spiral, a radial abundance gradient exists
\citep[e.g.,][]{zkh94a,vanzee98a}, with lower abundances at larger
galactocentric radius.  During the interaction, low-metallicity gas
from the galaxy outskirts is torqued into the high-metallicity galaxy
center \citep{mh96a,bh96a}, resulting in gas with a lower average
abundance (as first suggested in \citealt{kewley06b}).  We discussed
the quantitative details of this scenario for present-day equal-mass
mergers in \citet{rupke10a}.

A prediction of this model is that this gas redistribution
dramatically flattens the initial radial metallicity gradient very
shortly after first pericenter \citep{kewley06b,rupke10a}.
\citet{chien07a} found no oxygen gradient along the tidal tails of the
Mice, consistent with strong radial mixing of gas.  Shallow or broken
gradients are also present in many nearby barred galaxies, which are
undergoing radial redistribution of gas due to bar-induced gas motions
\citep{ve92a,zkh94a,mr95a,rw97a,dr99a}.  Barred galaxies also show
lower central metallicity than galaxies of similar morphological type
and luminosity \citep{dr99a}, as is observed in mergers.  Numerical
models of barred galaxies reproduce this behavior \citep{fbk94a}.

In a companion paper \citep{kewley10a}, we showed that eight
interacting galaxies which are underabundant with respect to the
luminosity-metallicity relation all have gradients that are flatter
than the nearby spirals M83, M101, and the Milky Way.  This result
strongly supports the model of merger-induced metal mixing.  In the
present paper, we expand on this result by giving a detailed analysis
of the systems in \citet{kewley10a}, doubling the sample of
interacting galaxies with measured gradients (to 16), and presenting a
comprehensive comparison to isolated spirals and the numerical
simulations.  The interacting galaxies in this sample are all in the
early stages of merging.  They are after first passage of the two (or
more) galaxies, but probably prior to second passage.  A future paper
will present similar data on galaxies in the later stages of merging
(Rupke et al. 2010, in prep.).

Understanding chemical evolution during a major merger will shed light
on galaxy evolution at the peak of cosmic star formation, since the
major merger rate, especially involving gas-rich galaxies, was higher
in the past \citep{deravel09a,bundy09a}.  These major mergers have
likely played an important role in the production and growth of
massive, red galaxies since $z\sim1$
\citep[e.g.,][]{bundy09a,robaina10a}.  Simulations suggest that dry
mergers (with no gas) flatten stellar metallicity gradients in
ellipticals \citep{dimatteo09a}.  It may be that wet mergers that form
bulge-dominated systems may do the same, if significant star formation
occurs after the gas gradient flattens.  In short, the metallicity
redistribution caused by gas-rich mergers may significantly impact
galaxy chemical evolution.

In \S\S\ 2 and 3, we present the observations and our analysis.  In
\S\ 4 we present the metal distributions, and then compare to
numerical simulations in \S\ 5.  We summarize in \S\ 6.  Cosmological
quantities are computed using $H_o = 73$~\kms, $\Omega_m = 0.27$, and
$\Omega = 1$, and in the rest frame defined by the cosmic microwave
background (where applicable).

\section{SAMPLE AND OBSERVATIONS} \label{sec:samp-and-obs}

\subsection{Sample} \label{sec:sample}

Our sample consists of galaxy pairs undergoing strong interactions in
the local universe.  The pairs are chosen from both optically-selected
\citep{arp66a,bgk00a} and infrared-selected \citep{sanders03a,ssm04a}
catalogs of interacting systems.  The primary selection criteria, to
ensure that the systems are strongly interacting, are that the
projected nuclear separation is small ($\la$ 30 kpc), the galaxy mass
ratios are small (1:1$-$1:3), and that the systems are obviously
disturbed (i.e., after first passage).  To distinguish from systems in
later stages of merging, we have rejected systems with (1) projected
pair separations $<$15~kpc and/or (2) compact morphology and faint
tidal features that indicate the system is near or after nuclear
coalescence.  The first criterion may remove some systems very near
first or second pericenter, but also securely rules out most systems
after second pericenter, since rapid orbital decay in strong,
equal-mass interactions leads to small nuclear separations after
second pericenter \citep[e.g.,][and references therein]{barnes96a}.

We have chosen spiral-spiral interactions to maximize the prevalence
of \htwo\ regions.  In each system in our sample, there is an
obviously interacting pair with small projected separation, near-equal
masses, and signs of interaction.  However, in some cases there is
also a third galaxy at large projected separations ($\ga$50~kpc) from
the dominant pair which may or may not be involved in the interaction.
Given their large distances, we do not further consider these possible
third members of the system.

Basic properties of individual galaxies in our sample are listed in
Table \ref{tab:sample}.  Our sample of 9 pairs/groups have fairly
average properties for nearby galaxies (Figure \ref{fig:sample}).
Individual galaxies are evenly distributed around $L_K^*$ ($\pm2$ mag;
\citealt{kochanek01a}), and have total infrared luminosities near
$L_{IR}^*$ \citep{sanders03a}.  As mentioned above, the pair mass
ratios (as traced by $K$-band luminosity) are in the 1:1$-$1:3 range,
with projected nuclear separations of $15-30$~kpc (except for Arp 248,
where the separation of the two primary galaxies is 54.3~kpc).

Table \ref{tab:sample} also lists optical diameters, inclinations, and
line-of-nodes position angles for the sample.  Many of these
parameters are taken directly from HyperLeda \citep{paturel03a}.  For
the NGC~2207 / IC~2163 system, we used parameters determined from
numerical models \citep{elmegreen95a,elmegreen00a}.  In the cases of a
few galaxies, the tidal stretching of the galaxies or general
interaction-induced disturbances make determining disk orientations a
challenge.  In this handful of cases, we have turned where possible to
\hone\ maps to determine the position angles.  These cases are listed
in the table notes.  We have also used this \hone\ data and inspection
of visual images to double-check HyperLeda inclinations.  The
inclinations of two galaxies remain uncertain: Arp 248 NED02 and Arp
256 NED02, which are particularly tidally-stretched.  In both of these
cases, the effect on the gradient is compensated for somewhat by a
larger measured \rtf.  We estimate that the inclinations lie in the
range $0-45^\circ$, and compute the deprojection using 30$^\circ$.

\subsection{Control} \label{sec:control}

A control sample of isolated galaxies is a necessary component of this
study.  In particular, we need to compare our results to a group of
galaxies that plausibly represent the galaxies in our sample as they
were before a major interaction.  We have selected our control sample
from published data on nearby, isolated spirals that have data on
\htwo\ region emission-line fluxes.  In particular, we have focused on
mid-to-late-type spirals with a mix of bar strengths
\citep{martin95a,dr99a}, since mid-to-late types are best represented
in the literature.  Gradients correlate with bar strength in a
relationship with significant scatter \citep{dr99a}, and depend weakly
or not at all on spiral type \citep{zkh94a}.  Our sample should thus
be representative of isolated spirals at the level allowed by small
number statistics.  Finally, with only two exceptions, we have picked
galaxies from moderate-size, homogeneous studies to minimize
systematic uncertainties.  (The two exceptions, NGC 300 and M101, are
well-characterized benchmarks in gradient studies, and so we include
them as well.)  The salient properties of our control sample, and the
references for the sample properties, are listed in Table
\ref{tab:control}.

\begin{figure}
  \plotone{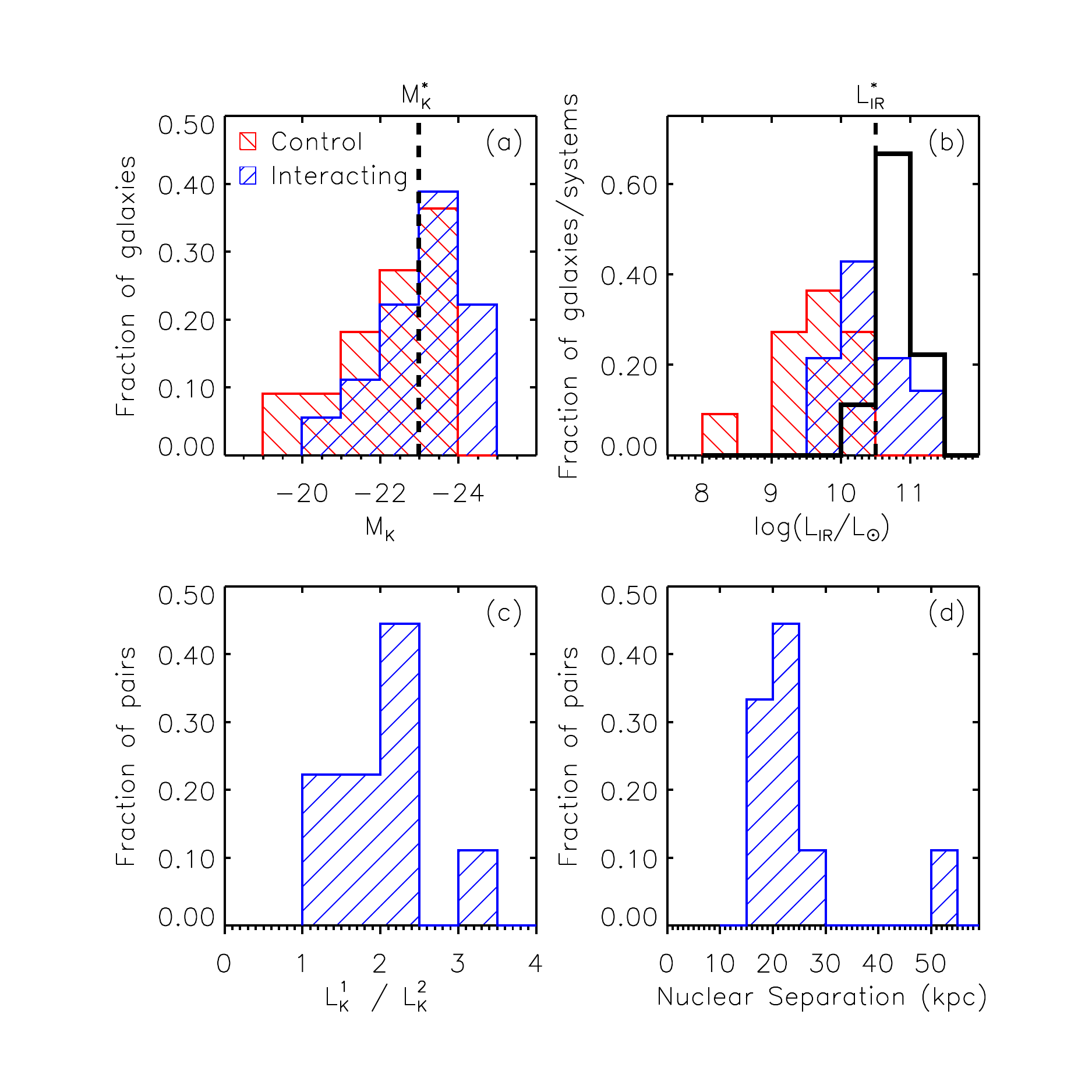}
  \caption{Properties of the control and interacting galaxies samples:
    (a) $K$-band absolute magnitude; (b) $8-1000$\micron\ infrared
    luminosity; (c) $K$-band luminosity ratio; and (d) nuclear
    separation.  The control and interacting samples are shown as red
    and blue hatched histograms.  In (b), the open histogram is the
    total system luminosity for the interacting sample.  In (c), the
    convention is such that the luminosity ratio is always $>$1.  The
    control and interacting samples are very similar in $M_K$, but the
    interacting galaxies have higher infrared luminosities.  The
    interacting galaxies have luminosities near $L^*$, mass ratios in
    the range 1:1$-$1:2.5, and (projected) nuclear separations in the
    range $15-30$ kpc.}
  \label{fig:sample}
\end{figure}

The interacting and control samples are well-matched in mass, as
traced by $K$-band luminosity (Fig.~\ref{fig:sample}$a$; the control
sample is only half a magnitude fainter on average).  They are also
well-matched in optical radius, with average (median) optical radii of
\rtf\ $=$ 13 (12) kpc in both samples.  However, the interacting
galaxies have an average (median) infrared luminosity that is boosted
a factor of 7 (4) above the control sample
(Fig.~\ref{fig:sample}$b$).  This is unsurprising, given the expected
higher star formation rates in the interacting sample \citep{bgk00a}.

Rather than use gradient determinations published in the literature,
we have used the published line fluxes to measure abundances using the
same strong-line diagnostic that we apply to our interacting galaxies
sample (\S\ref{sec:spectral-analysis}).  We have also performed our
own spatial deprojections using published photometric parameters (see
references in Table~\ref{tab:control}).  The resulting oxygen
abundance gradients, in terms of dex/kpc and dex/\rtf, are given in
Table \ref{tab:control}.

\subsection{Observations} \label{sec:observations}

We used the multiobject spectroscopic capabilities of the Keck
Low-Resolution Imaging Spectrometer (LRIS;
\citealt{oke95a,mccarthy98a}) to observe \htwo\ regions in the
galaxies of interest.  The excellent blue sensitivity of LRIS was
important for measuring the \otl\ emission lines at high S/N; the \ot\
doublet is used in our strong-line metallicity diagnostics
(\S\ref{sec:data-reduct-analys}).  Our typical setup was to observe
the wavelength range $3500-7000$~\AA, using the 900 l/mm grating in
the red and either the 400 or the 600 l/mm grism in the blue, and the
560~nm dichroic to separate the red and blue spectra.  (For one
galaxy, Arp 248, we were forced to use the 300 l/mm grism in the blue
to obtain the entire spectrum, because LRIS-R was inoperable.)  For
over half of our data, the LRIS atmospheric dispersion corrector (ADC)
was in place, mitigating any light loss in the blue.  For the other
half (taken prior to mid-2007), slit losses due to atmospheric
refraction were generally small.  Total exposure times per mask were
typically an hour.

Slitmasks were created in most cases using sensitive \ha\
emission-line images.  Slits were 1\arcsec\ wide.  The \ha\ images
were taken using redshifted narrowband filters at either the Vatican
Advanced Technology Telescope (VATT; E. Barton and R. Jansen 2005,
private comm.) or the University of Hawaii 88-inch telescope.
However, in two cases (Arp 256 and Arp 298), the spectroscopic data
was taken in the context of a program to observe star clusters (Chien
et al. 2010, in prep.).  In this case, star clusters were targeted
using {\it Hubble Space Telescope} ($HST$) broadband F435W images.
Since significant H$\alpha$ emission fell in the slits in these cases,
we were able to extract useful \htwo\ region spectra for the current
work.

\section{DATA REDUCTION AND ANALYSIS} \label{sec:data-reduct-analys}

\subsection{Data Reduction} \label{sec:data-reduction}

The LRIS data were reduced using a pipeline we developed specifically
for the current dataset.  This pipeline combines Image Reduction and
Analysis Facility (IRAF) and Interactive Data Language (IDL) scripts.
For each side of the spectrograph, the steps we took were as follows:
bias subtraction; slit identification and tracing using flatfield
exposures; arc lamp identification; slit extraction and spatial
rectification (including any atmospheric dispersion correction, if
required); a small shift to match dispersion solutions among
exposures; data combination, including noise-threshold cosmic ray
rejection; wavelength calibration; extraction of spectra; sky
subtraction; and flux calibration.  In the slit identification step,
slit edges were detected using a directional filter that computed the
first derivative of the flatfield in the spatial direction.  We
extracted each local peak in H$\alpha$ using a
1\arcsec$\times$1\arcsec\ aperture, and subtracted the sky using
off-object spectra free of galaxy continuum and line emission.  Sky
subtraction was generally imperfect due to the change of the LRIS
point spread function across the $6\arcmin\times8$\arcmin\ field-of-view
combined with the use of a small number of sky spectra that were
typically on the field edges (to avoid galaxy contamination within the
on-source slits).

Blue and red spectra were reduced separately, and then combined near
5600~\AA\ by multiplicatively matching the continuum fluxes in this
region.  Following the analysis of the spectra, this relative flux
calibration was further refined.

Spectra were mapped to positions on the sky assuming correct alignment
of each slitmask during observations along with knowledge of the slit
plane to focal plane mapping.  (We used at least four alignment stars
per slitmask to ensure correct mask alignment.)  Galactocentric radii
were then computed using information on the 3-dimensional alignment of
each galaxy (\S\ref{sec:sample}).  To determine the center of all
galaxies except NGC 3994/5, we used near-infrared positions from
2MASS.  Sloan Digital Sky Survey (SDSS) centers were used for NGC
3994/5.  Though these systems are gravitationally disturbed, the
near-infrared is able to probe the central bulge in each galaxy and
thus yield a reasonably accurate true gravitational center.

\subsection{Spectral Analysis} \label{sec:spectral-analysis}

We extracted information from the resulting spectra by performing
detailed fits to the stellar continuum and line emission.  The
software fitting package that we wrote for this purpose, UHSPECFIT,
was based on the stellar fitting routines of \citet{mk06a}, who fit
stellar continua using a linear combination of several template
spectra.  We built on these routines a sophisticated suite of
emission-line fitting software that fits all emission lines
simultaneously and allows for multiple velocity components, starting
with the code framework used in \citet{zahid10a}.

To fit these relatively high-resolution data, we used the stellar
population synthesis models of \citet{gonzalezdelgado05a} as our
stellar continuum templates.  We relied on the solar metallicity,
Geneva isochrone models for our fits.  The quality of the fits is not
a sensitive function of stellar metallicity; since our goal was not to
constrain the stellar metallicity, but rather to remove the stellar
continuum, we used only the solar metallicity models.  The stellar
fits were performed on the blue half of the spectra (from $3700-5500$
\AA\ in the rest frame) after removal of wavelength regions near
strong emission lines; the resulting stellar fit was then normalized
to fit the red half of the spectra.  We subtracted the stellar
continuum from the spectra and performed emission-line fits.  Two
example fits are shown in Figure \ref{fig:example_fit}.  One is a high
emission line flux case, and the other is a low emission line flux
case (roughly 1$\%$, in \ha\ flux, of the high flux case).  These are
illustrative of the typical high fidelity of the fits, and reveal that
fits to the Balmer absorption lines in particular are very good.

Once the fits were complete, a final blue-red relative flux
calibration was performed by comparing the measured reddening from the
\hb/\hg\ emission line ratio (using the extinction law of
\citealt{cardelli89a}) with that measured from the \ha/\hb\ ratio.
The discrepancy was used to correct the blue emission-line fluxes.
This method worked very well for our data (which showed high
signal-to-noise in \hg) and our choice of spectral fitting (which
permitted robust separation of Balmer absorption and emission).
However, our abundance errors increase with the use of \hb/\hg\ to
determine the reddening (as opposed to \ha/\hb).

We corrected the emission line fluxes for reddening
\citep{cardelli89a} and computed important line ratios.  Spectra with
very uncertain reddening because of poorly-fit or weak \hb\ and \hg\
emission lines were ignored.  The dereddened line ratios were then
used to classify each spectrum as \htwo-region, composite, Seyfert, or
LINER \citep{kewley06a}.  We calculated abundances only for those
regions which are securely of \htwo-region origin, without probable
contamination from other sources of excitation.  We included only
those data identified as \htwo-region spectra based on the \ntl/\ha\
vs. \othl/\hb\ diagnostic (i.e., regions that fall below the
\citealt{kauffmann03a} star formation locus) and \htwo\ or composite
spectra based on the \ool/\ha\ vs. \othl/\hb\ diagnostic (i.e.,
regions that fall below the \citealt{kewley01a} extreme starburst
line).  For both line ratio cuts, we included data up to 0.1~dex to
the right of the diagnostic line to account for possible
uncertainties.

Roughly 25\% of our spectra were rejected because of poorly-fit \hg\
emission, line ratio cuts, or inadequate line fits.  Spectra rejected
due to line ratio cuts are shown as black crosses atop the galaxy
images in Figure \ref{fig:results}.  Our final sample of \htwo\
regions in early-stage interacting galaxies consists of 332 spectra.
These are accompanied by 281 \htwo\ region spectra from the control
sample.  Four nuclei in our sample show either Seyfert (NGC 2207, IC
2163, NGC 7469) or LINER (NGC 3994) spectra based on our data.
However, these nuclear ionizing sources have minimal impact on the
extra-nuclear \htwo\ regions that are used to compute abundances;
phenomena like ionization cones would in any case be detected and
rejected using our line ratio cuts.

In this paper, we quote abundances calculated from the \ntl/\otl\
abundance diagnostic \citep[][hereafter KD02]{kd02a}.  Of the
strong-line diagnostics, this one produces the least root-mean-square
(RMS) dispersion about the mean due to its relative insensitivity to
ionization parameter (KD02).  To study possible systematic errors from
the use of this diagnostic (due to, e.g., improper extinction
corrections or improper relative flux calibration of the blue and red
LRIS arms), we computed abundances using the \citet[][hereafter
KK04]{kk04a} \rtt\ and \citet[][hereafter PP04]{pp04a}
(\oth/\hb)/(\nt/\ha) diagnostics (following the guidelines laid out by
\citealt{ke08a}) and then converted these diagnostics into the KD02
calibration using the formulas of \citet{ke08a}.  The results are
shown for all of the \htwo\ regions in this paper in Figure
\ref{fig:abundcomp}, and for individual systems in Figure
\ref{fig:results}.

It is clear from examining Figure \ref{fig:abundcomp} that, overall,
there are no significant systematic discrepancies introduced by our
data reduction and analysis that make our \nt/\ot\ line ratios
suspect.  For instance, both KK04 and PP04 are less sensitive to
extinction than KD02, and inaccurate extinction corrections should
show up as deviations from equality in these figures.  The interacting
galaxies sample does in fact show a 0.05~dex offset from equality in
the KD02 vs. PP04 diagram.  However, the PP04 diagnostic does not
correct for ionization parameter variations, which introduces a
systematic uncertainty in this diagnostic; the KD02 diagnostic is
insensitive to ionization parameter.  Furthermore, the offset is well
within the expected uncertainties of the diagnostic and conversion
formulas, and as a simple offset, will have no impact on relative
abundances like gradients.  In fact, the RMS dispersion about the line
of equality in the interacting galaxies sample (0.05~dex) is
significantly lower than that in the control sample for the KD02
vs. KK04 diagram, suggesting that our data is highly uniform.  In the
KD02 vs. PP04 diagram, the RMS dispersions are the same for the two
samples, but for the interacting sample this is in large part due to
the systematic offset to one side of equality.

Errors in the emission-line fluxes were computed by assigning the
error in the peak flux to equal the RMS residual within 2$\sigma$ of
each line center, after subtracting the continuum and line fits from
the data.  Errors were propagated primarily using analytic
expressions, but for abundance and gradient errors we employed Monte
Carlo methods.

In the following section, we use the calculated abundances and radii
for the \htwo\ regions in each galaxy to compute radial abundance
gradients, and compare the gradients of the control and interacting
samples.

\section{OXYGEN GRADIENTS} \label{sec:oxygen-gradients}

With abundances and deprojected radii in hand, we computed radial
oxygen gradients for each galaxy in our sample.  In two cases (IC~5283
and UGC 12915), the radial coverage and number of \htwo\ regions
available were each too small to yield a reliable estimate of the
gradient.  In the control sample, we limited the gradient fits to
within 1.5\rtf, to match the radii fit in the interacting galaxy
sample.  We wished to avoid biasing the gradient fit toward particular
locations in the galaxy (e.g., to high-luminosity \htwo\ regions), so
we performed unweighted least-squares fits.  Our fits were computed
with radius as the independent variable, and errors in slope and
intercept were estimated using Monte Carlo methods.  The resulting
gradients and intercepts are given in Table \ref{tab:gradients}.

In Figure \ref{fig:results}, we present for each system the results of
our emission-line analysis: emission line diagnostic diagrams,
comparison of different metallicity diagnostics, a two-dimensional
projected oxygen abundance map, and a one-dimensional deprojected
radial oxygen abundance profile.

The measured gradients in these interacting systems are on average
significantly flatter than those in the control sample (Figure
\ref{fig:hist-grads}).  The control sample exhibits median slopes of
-0.041$\pm$0.009 dex/kpc and -0.57$\pm$0.05 dex/\rtf, while the
interacting systems have median slopes of -0.017$\pm$0.002 dex/kpc and
-0.23$\pm$0.03 dex/\rtf\ (where the errors given are the standard
error).  In short, the interacting systems have gradients that are, on
average, less than half as steep as isolated galaxies.

In \citet{kewley10a}, we showed that gradients in eight of the systems
presented here were shallower than gradients in M83, M101, and the
Milky Way.  The present data quantify this difference at a high
confidence level using large control and interacting samples.  Taken
by themselves, these differences in gradients are a strong
confirmation of the model of merger-induced gas inflow leading to
metal mixing.

Surprisingly, the dispersions in the interacting sample gradient
distributions are also less than the dispersions in the control
sample: the control sample shows standard deviations of 0.03 dex/kpc
and 0.17 dex/\rtf, compared to 0.01 dex/kpc and 0.11 dex/\rtf.  The
explanation for the latter effect is unclear, but could result from
the origin of the \htwo\ region data in heterogeneous and homogeneous
datasets, respectively.

No significant correlations are seen between gradients and galaxy near
infrared and total infrared luminosities, either in the interacting
sample or the control sample (Figure \ref{fig:grads-vs-gal}).
Previous work has shown that correlations may exist between optical
luminosities and gradients in dex/kpc \citep{zkh94a}; however, such a
trend apparently does not persist in interacting systems (and is not
obvious in our control sample).  One might expect a correlation of
infrared luminosity with gradient in interacting systems, if \lir\ is
assumed to trace star formation rate and star formation is driven by
gas inflow.  Again, no such trend is evident, though we only have
upper limits in \lir\ for much of our sample.

No obvious trends are seen when comparing gradients to system
properties that may parameterize the stage or strength of the
interaction: projected nuclear separation, near-infrared luminosity
ratio, or total system infrared luminosity (Figure
\ref{fig:grads-vs-sys}).  Projected nuclear separation is of course
only approximate, since we cannot deproject the two galaxies without
detailed modeling, and there is a degeneracy in the age of the
interaction between galaxies moving apart after first pericenter and
returning to second pericenter.  Furthermore, mass ratios in our
sample (as traced by $K$-band luminosity ratio) are all close to
unity, and one might expect little variation among the systems in our
sample based on this parameter alone.  Finally, confusing the matter,
simulations also show that although both galaxies in a strongly
interacting pair experience a flattening of the gradient very quickly
after first pericenter, the relative degree of flattening of the two
disks varies depending on the geometry and the distance of first
pericenter \citep{rupke10a}.  This effect would introduce scatter into
any underlying correlation between gradients and system properties.
See \S\ref{sec:gradients-vs.-merger} for further discussion.

Some systematic uncertainty arises in our abundance gradients due to
our choice of abundance calibration/diagnostic.  Some calibrations
(including both strong- and weak-line ones) may yield shallower slopes
than we calculate \citep[e.g.,][]{ke08a,bresolin09b}.  However,
\citet{kewley10a} showed that an \rtt-based diagnostic yields the same
results as the KD02 diagnostic we employ in this paper.  More
importantly, we rely on the fact that within a particular abundance
calibration/diagnostic, {\it relative} abundances are reliably
computed and internally consistent \citep{ke08a}.  When using detailed
chemodynamical models to constrain parameters of individual systems,
measuring the absolute value of the gradient precisely may be
important.  Within the context of the current study, we are concerned
only with gradient changes relative to initial conditions.  We also
emphasize that, were the initial galaxy gradients too shallow (or
completely flat), the metal dilution clearly observed in interacting
systems would be impossible, and would thus be inconsistent with
empirical evidence.

Nevertheless, as a check on the reliability of our \nt/\ot\ gradients,
Figure~\ref{fig:results} shows, for each system, how the \nt/\ot\
abundances compare with those computed in the KK04 and PP04
diagnostics (see \S\ref{sec:spectral-analysis} for more details).  As
is seen in the full sample (Fig. \ref{fig:abundcomp}), individual
systems are largely free of systematic discrepancies.  The two
exceptions are NGC~2207/IC~2163 and Arp 248, which show discrepancies
between the KD02- and PP04-computed metallicities.  These anomalies
may result from variations in ionization parameter that remain
uncorrected in the PP04 diagnostic (\S\ref{sec:spectral-analysis}) or
reddening uncertainties.  Further analysis shows that, if we were to
use the PP04 abundances (after being converted into the KD02 system),
we would measure shallower gradients in Arp 248 NED01 and Arp 248
NED02 and a steeper gradient in NGC 2207 (but not IC 2163), in each
case by a factor of $\sim$2.

With the common exception of the galaxy nuclei, the vast majority of
data in our sample are pure \htwo\ regions.  However, UGC 12914/12915
has a significant number of regions indicative of non-photoionized
emission.  In fact, the north half of UGC 12914 has a significant
number of regions with multiple velocity components and shock-like
line ratios \citep{allen08a,rich10a}.  This region overlaps a bridge
of \hone\ gas between the galaxies \citep{condon93a}.

\section{COMPARISON TO SIMULATIONS} \label{sec:comp-simul}

Numerical simulations are a powerful tool for investigating the
evolution of galaxy mergers.  However, most simulations of mergers to
date have paid little attention to gas-phase metallicity and its
evolution.  Recently, \citet{rupke10a} discussed metallicity evolution
prior to second pericenter in smoothed particle hydrodynamic (SPH) /
N-body simulations of equal-mass mergers with gas mass fractions of
10\%\ (to match nearby galaxies).  Though they did not include
enrichment from ongoing star formation, \citet{rupke10a} reached good
agreement with the magnitude of deviations of major mergers from the
mass-metallicity relation.  They also showed that radial abundance
gradients should flatten rapidly after first pericenter, and predict
the relationship between gradient flattening and nuclear abundance and
the detailed shape of the gradient with time.  \citet{montuori10a}
presented simulations including star formation that largely agreed
with \citet{rupke10a}, but did not address the evolution of gradients.
Torrey et al. (2010, in prep.) find similar results to
\citet{montuori10a}, but show that preferential consumption of
high-metallicity gas by star formation in galaxy nuclei plays a role
alongside metal dilution due to inflow.

In the following sections, we compare our data on metallicity
distributions in interacting galaxies to the simulations of
\citet{rupke10a}.  These SPH/N-body simulations are described in
detail in \citet{barnes04a}.  In brief, 24576 gas particles were
distributed in an exponential disk of scale height equal to 6\%\ of
the disk scale length.  The stars occupied a bulge and identical disk,
with a bulge-to-disk ratio of $1/3$.  Eight close-passage mergers were
simulated, with all combinations of direct and retrograde considered.
The metallicity per gas particle was unchanged during the simulation,
and two different initial metallicity gradients were considered (-0.2
dex/$R_{disk}$ and -0.4 dex/$R_d$).  In this work we modified the
initial gradients slightly to match our control sample: -0.15
dex/$R_d$.  This number results from the typical gradient in dex/\rtf\
for the control sample (\S\ref{sec:oxygen-gradients}), when converted
to dex/$R_d$ using a typical ratio for \rtf/$R_d$
(\S\ref{sec:grad-lumin-metall}).

\subsection{Gradient Distribution} \label{sec:grad-distr}

Our galaxy sample was chosen to represent major mergers somewhere
between first and second pericenter.  The distribution of observed
gradients is shown in Figure \ref{fig:hist-grads}.  For comparison, we
construct a predicted gradient distribution from our set of simulated
major mergers, assuming that our sample is randomly distributed
between first and second pericenter.  We also assumed that the
simulated galaxies start with a single gradient equal to the mean
value in the control sample.  The observed and simulated distribution
show remarkable agreement (Figure \ref{fig:hist-grads-sims}).

\subsection{Gradients vs. Merger Stage} \label{sec:gradients-vs.-merger}

We show in Figure \ref{fig:grads-vs-sys} that, for the present sample,
there is no strong dependence of gradient on merger stage, as
parameterized by projected nuclear separation.  This is consistent
with the predictions of our simulations.  Figure
\ref{fig:grads-vs-nsep-sims} reveals that, between first and second
pericenter, there is no predicted dependence of the gradients on
actual nuclear separation, for the ensemble of simulations as a whole.
Within particular mergers, a more recognizable dependence of gradient
on nuclear separation may emerge, but we must consider our present
dataset as random snapshots of an ensemble of different mergers.  We
do not at present have an accurate understanding of the initial
conditions of most or all of the systems in our sample.

\subsection{Gradients and the Luminosity-Metallicity
  Relation} \label{sec:grad-lumin-metall}

The motivation for the present work was the observation that
interacting galaxies fall below the luminosity-metallicity (\lz) and
mass-metallicity (\mz) relations of star-forming galaxies.  Do the
current data present a consistent picture in this regard, such that
the nuclear metallicities of interacting systems with shallow
gradients also fall below these relationships?  For a subset of the
current sample, we used the $B$-band \lz\ relation in
\citet{kewley10a} and found no correlation between gradients and \lz\
offsets, but were limited by small sample size and the fact that
$B$-band luminosity is sensitive to ongoing star formation and
extinction effects.

In this paper, we compare the control and interacting samples to the
luminosity-metallicity relation as measured in the near-infrared
\citep{salzer05a}, and use simulations to interpret the result.
Multicolor broadband photometry is not available for our entire
sample, so we are unable to measure accurate masses and apply the \mz\
relation.  Near-infrared photometry, however, is less sensitive to
extinction or contamination from star formation than the optical; it
is also a reasonable tracer of stellar mass within a factor of
$\sim$2 \citep{maraston05a}.

The \citet{salzer05a} abundances are available in several diagnostics.
None of these are the \nt/\ot\ diagnostic that we employ, so we resort
to the abundance conversion formulas of \citet{ke08a}.  We start with
the \rtt-computed abundances from \citet{salzer05a}; the diagnostic
they use is a functional fit of \rtt\ to abundances from modeling of
SDSS spectra \citep{cl01a,tremonti04a}.  We then apply the formula
from \citet{ke08a}.  Because two transformations are involved, we
consider these abundances to be only approximate, and are useful
primarily for establishing the slope of the \lz\ relation.

We do not have nuclear abundances measured from long-slit spectra for
all of the galaxies in the current sample.  We thus estimate
abundances by extrapolating our gradient fits, and compute the
abundances at a discrete radius of $0.1$\rtf.  This roughly matches
the extent of the long-slit apertures of the \citet{salzer05a} data,
which have $\langle z \rangle = 0.063$.

Using this method, neither the control nor interacting samples fall on
the $K$-band \lz\ relation.  This is presumably due to one of two
effects: (1) the \citet{salzer05a} abundances are luminosity-weighted
measurements from nuclear spectra, while ours are measured at a
fiducial radius using fits to \ion{H}{2} region data; and/or (2) the
uncertainties introduced by the two transformations involved in
computing the abundances of the \citet{salzer05a} sample.  The latter
effect is not unusual when converting among different metallicity
calibrations \citep{ke08a}.  To account for this effect, we apply a
constant offset of $-$0.3~dex to both the control sample and
interacting sample metallicities.  This offset minimizes the RMS
deviation of the control sample from the \lz\ relation.

The exact vertical normalization of the data is irrelevant, however;
the important quantities are the {\it relative} nuclear abundances of
the control and interacting samples at a given luminosity, which are
secure since we use the same abundance diagnostic and compute the
nuclear abundances in the same way.  The \citet{salzer05a} data simply
provide a more accurate slope for the \lz\ relation than the control
sample by itself.

The result of this exercise is displayed in Figure \ref{fig:lz-0p1}.
This figure makes evident that the early-stage interacting galaxies,
which have shallower gradients than the control sample, also fall
below the \lz\ relation defined by the control sample and emission
line galaxy sample.  In fact, the interacting galaxies form a
remarkably coherent \lz\ relationship that is simply offset from that
of isolated star-formers by $-$0.2~dex in abundance.

To tie together abundance offsets and gradients, we compare directly
to the simulations of \citet{rupke10a}.  These simulations predicted
that the gradient slope would change quickly after first pericenter,
while the nuclear metallicity would change more slowly between first
and second pericenter.  The simulations yield gradients in units of
exponential disk scale length of the gas disk (dex/R$_d$), while we
express observed gradients in physical units (dex/kpc) and optical
(stellar) isophotal radius (dex/\rtf).  For our control sample, we
have stellar disk scale lengths as well as optical radii, and the two
are related according to \rtf\ $=(3.7\pm1.2)$R$_d$, where the
uncertainty given is the standard deviation.  Using this scaling, we
compute the gradients in dex/R$_d$ for our interacting galaxy sample.

For the current data set, we measured abundances at a fiducial radius
of 0.1\rtf.  To match this procedure for the simulations, we computed
average abundances within a narrow, 0.1R$_d$-wide radial bin centered
on 0.35R$_d$.  (This differs from the method in \citealt{rupke10a}, in
which we averaged over the entire central disk within 0.5R$_d$.)  For
the data, we assume that the interacting galaxies started on the \lz\
relation of the control sample and have had their central metallicity
lowered during the merger.

It is clear that, given the uncertainties inherent in this procedure
(e.g., the galaxies didn't necessarily start on the mean \lz\
relation) and the simplicity of the simulations (which do not include
ongoing star formation), there is remarkable agreement between the
data and simulations (Figures \ref{fig:dz-vs-grad} and
\ref{fig:dz-vs-grad-lab}).  A total of $70-80\%$ of the interacting
galaxies fall directly in the phase space delineated by the simulated
evolutionary tracks.  Furthermore, the region of overlap with the
simulations is near the time of first turnaround (between first and
second pericenter) for many of the simulated pairs, which is
consistent with the galaxies in our sample being somewhere between
first and second pericenter on average.  (In a random distribution of
post-pericenter pairs selected by nuclear separation, half should be
pre-turnaround, and half post-turnaround.)

The data are thus quantitatively consistent with the prediction of the
simulations of \citet{rupke10a} that the gradient should flatten
quickly after first pericenter, and that the nuclear abundance should
decrease at a slower rate between first and second pericenter.

The scatter of plausible initial conditions, as represented by the
control sample, do caution against drawing detailed conclusions from
this comparison of data and simulations.  In particular, a variety of
initial conditions can explain the current position of the interacting
galaxies in the metallicity-gradient phase space.  The metallicity
scatter in the control sample is partly a measurement issue; e.g., the
scatter in the simulated control \lz\ relation is larger than in the
\citet{salzer05a} sample, and the control sample \htwo\ regions show a
larger RMS dispersion in abundance than the interacting sample
(Fig.~\ref{fig:hist-rms}).  To deal with this metallicity scatter,
whatever its origin, however, we simply shift the simulations left or
right in Figure \ref{fig:dz-vs-grad} to start at the control sample
initial conditions.

A shift in \lz\ offset could in principle resolve any discrepancies
between data and simulations for individual galaxies.  For example,
NGC 2207 is clearly not at a late merger stage, as suggested by its
position with respect to the merger models (it is probably just after
first passage; \citealt{elmegreen95a}).  However, it could initially
have had lower metallicity than predicted by the mean \lz\ relation
and/or a steeper initial gradient, leading to a more favorable
comparison with models in Figure \ref{fig:dz-vs-grad}.

If a galaxy were to start with a larger gradient than is assumed in
the models, then the changes in gradient and central metallicity are
larger \citep{rupke10a}.  For instance, if we assume the initial
conditions represented by NGC~300 or NGC~2997, the simulations still
overlap the data, due to the more extreme changes that result from a
steeper gradient and the same amount of infall (Figure
\ref{fig:dz-vs-grad-steep}).  Closer inspection rules out this
scenario for most systems, however, since (a) the overlap of the data
and the bulk of the simulations is not as good and (b) these initial
conditions would require almost all of the observed interacting
systems to lie near second pericenter.  This is unlikely, due to how
we have selected our sample (\S\ref{sec:sample}).  The
simulation-to-data comparison shown in Figure \ref{fig:dz-vs-grad},
with half the initial gradient of that in NGC~300 or NGC~2997, is
therefore a more likely scenario for most systems.

\subsection{Gradient Profiles}

Following standard practice, we have characterized radial metallicity
profiles using straight line fits [i.e., 12$+$log(O/H) $=$ a $+$ bR].
However, the predicted metallicity profiles from the simulations of
\citet{rupke10a} do show distinctive shapes that are not purely
straight lines at all times.

Individual galaxies in our sample do not have enough data points to
accurately characterize the profile shapes beyond straight lines.
However, we can compute the average deviation from straightness as a
function of radius by subtracting the gradients and binning the data
in radius.  We do this for both the control and interacting samples,
and show the results in Figure \ref{fig:profile}.  These two samples
show metallicity profile residuals that are significantly different.
The control sample is largely consistent with a straight line.
However, the interacting sample shows a clear ``concave up'' shape, in
the sense that positive deviations are seen at small and large radius,
with negative deviations at middle radii.

How does this compare with simulations?  To bin the simulation data,
we consider the ensemble of profiles at first and second pericenter
and first turnaround.  The result, with the interacting galaxies data
overplotted, is shown in Figure \ref{fig:profile}.  It is apparent
that for the simulations, there is a small upturn at low radius at all
times, consistent with the data.  Though there is not exact
quantitative agreement in any of the three stages, the simulations at
first turnaround show the same ``concave up'' shape seen in the data.
By contrast, the simulations at 1st and 2nd pericenter show a flatter
profile outside the nuclear regions.  As discussed above, this is
consistent with the idea that, on average, our sample is near first
turnaround, and otherwise evenly distributed between first and second
pericenter.

Given the complexity and variety of the actual merging systems in this
study, it is impressive that there remains such good agreement at this
level of comparison between data and simulations, and speaks to the
robustness of the dilution model and its predictions.  It also
suggests that metals produced in star formation initiated by the
merger have only a small effect on the gas-phase metallicity
distributions prior to second pericenter.  This could occur either
because most star formation occurs after second pericenter or because
of inefficient mixing of the metal-enriched hot phase with the cooler
ISM.  In fact, only a limited fraction of the gas in interacting
galaxies can be consumed in star formation prior to merging, in order
to leave a significant gas reservoir at late stages \citep{sss91a}.

\section{SUMMARY AND PROSPECTS} \label{sec:summary-prospects}

We have presented here data on the gas-phase oxygen abundance
gradients in 9 pairs/groups (18 galaxies) in the early stages of an
interaction (after first pericenter, prior to second).  We computed
radial gradients based on \htwo\ region spectroscopy and the most
precise strong-line abundance diagnostic available (KD02).  We find a
clear break between the gradient distributions of isolated and
interacting systems: interacting systems consistently have gradients
about half that in normal, isolated spirals.

We compare the data to the simulations of \citet{rupke10a}, and find
remarkable statistical agreement at the ensemble level.  The
simulations accurately reproduce (1) the gradient distribution, (2)
the lack of gradient dependence on projected nuclear separation, (3)
the relationship between gradients and deviations from the
luminosity-metallicity relation, and (4) metallicity profile
deviations from a straight line.

These data are clear evidence that models of metal redistribution due
to interaction-induced gas inflow
\citep{kewley06b,rupke10a,montuori10a} are able to account for the low
nuclear abundances observed in interacting galaxies
\citep{lee04a,kewley06b,rvb08a,ellison08a,michel08a,peeples09a,ekta10a,solalonso10a}.
Now that this mechanism is firmly established, future work should be
devoted to understanding how this affects the global evolution of the
mass-metallicity relation over cosmic times and the formation of
elliptical galaxies.  In particular, if mergers or post-mergers
dominate samples of galaxies used to determine the $\mz$ relation at
high redshift, this bias must be taken into account.  Furthermore, all
massive galaxies have experienced mergers during their lifetime.  The
re-distribution of metals due to gas inflow will then affect the
stellar populations formed at later times at all radii.

Though we have already found remarkable agreement between the data and
simulations at a statistical level, this comparison could be improved
upon using simulations that incorporate realistic models of star
formation and chemical enrichment (including a sophisticated model of
starburst-driven galactic winds).  Secondly, we have compared to
simulations only after binning the data azimuthally; more detailed
comparison of theory to actual projected metallicity distributions
would provide a stronger understanding of how and where the dilution
occurs.  Furthermore, we have only looked for agreement at a
statistical level.  Though the current data already provide broad
constraints on the simulations, more detailed constraints will require
careful modeling of individual systems.

The limited merger ages traced by this early-stage sample constrain
our ability to trace metal redistribution over the entire course of a
merger.  In future work (Rupke et al. 2010, in prep.), we will present
data on later stage mergers to accomplish this.

\acknowledgments

The authors thank Jabran Zahid for sharing the software that formed
the basis of UHSPECFIT; Fabio Bresolin for data analysis tips, a
galaxy deprojection algorithm, and commenting on the manuscript;
Margaret Geller for commenting on the manuscript; Betsy Barton and
Rolf Jansen for providing their H$\alpha$ images; Josh Barnes for
helpful discussion; and the referee for useful comments regarding
content and presentation.  The data presented herein were obtained at
the W.M. Keck Observatory, which is operated as a scientific
partnership among the California Institute of Technology, the
University of California and the National Aeronautics and Space
Administration. The Observatory was made possible by the generous
financial support of the W.M. Keck Foundation.  The authors also
recognize the very significant cultural and spiritual role that the
summit of Mauna Kea has within the indigenous Hawaiian community.  We
are most fortunate to have the opportunity to observe the universe
from this mountain.

\def\eprinttmppp@#1arXiv:@{#1}
\providecommand{\arxivlink[1]}{\href{http://arxiv.org/abs/#1}{arXiv:#1}}
\def\eprinttmp@#1arXiv:#2 [#3]#4@{\ifthenelse{\equal{#3}{x}}{\ifthenelse{
\equal{#1}{}}{\arxivlink{\eprinttmppp@#2@}}{\arxivlink{#1}}}{\arxivlink{#2}
  [#3]}}
\providecommand{\eprintlink}[1]{\eprinttmp@#1arXiv: [x]@}
\renewcommand{\eprint}[1]{\eprintlink{#1}}
\providecommand{\adsurl}[1]{\href{#1}{ADS}}
\renewcommand{\bibinfo}[2]{\ifthenelse{\equal{#1}{isbn}}{\href{http://cosmolog%
ist.info/ISBN/#2}{#2}}{#2}}


\clearpage

\begin{figure}
  \centering
  \begin{tabular}{c}
    \epsfig{file=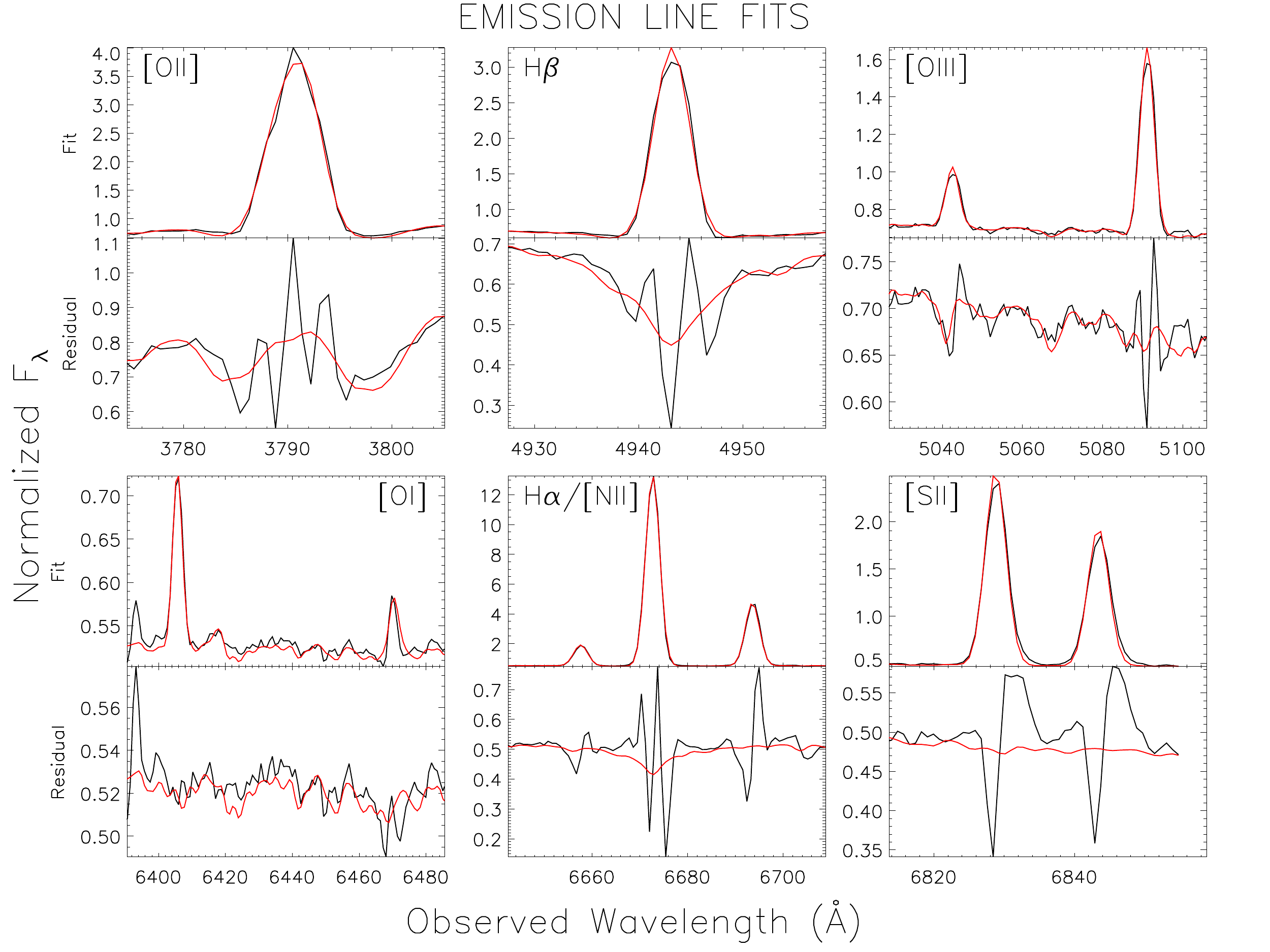,width=0.85\linewidth,clip=} \\
    \epsfig{file=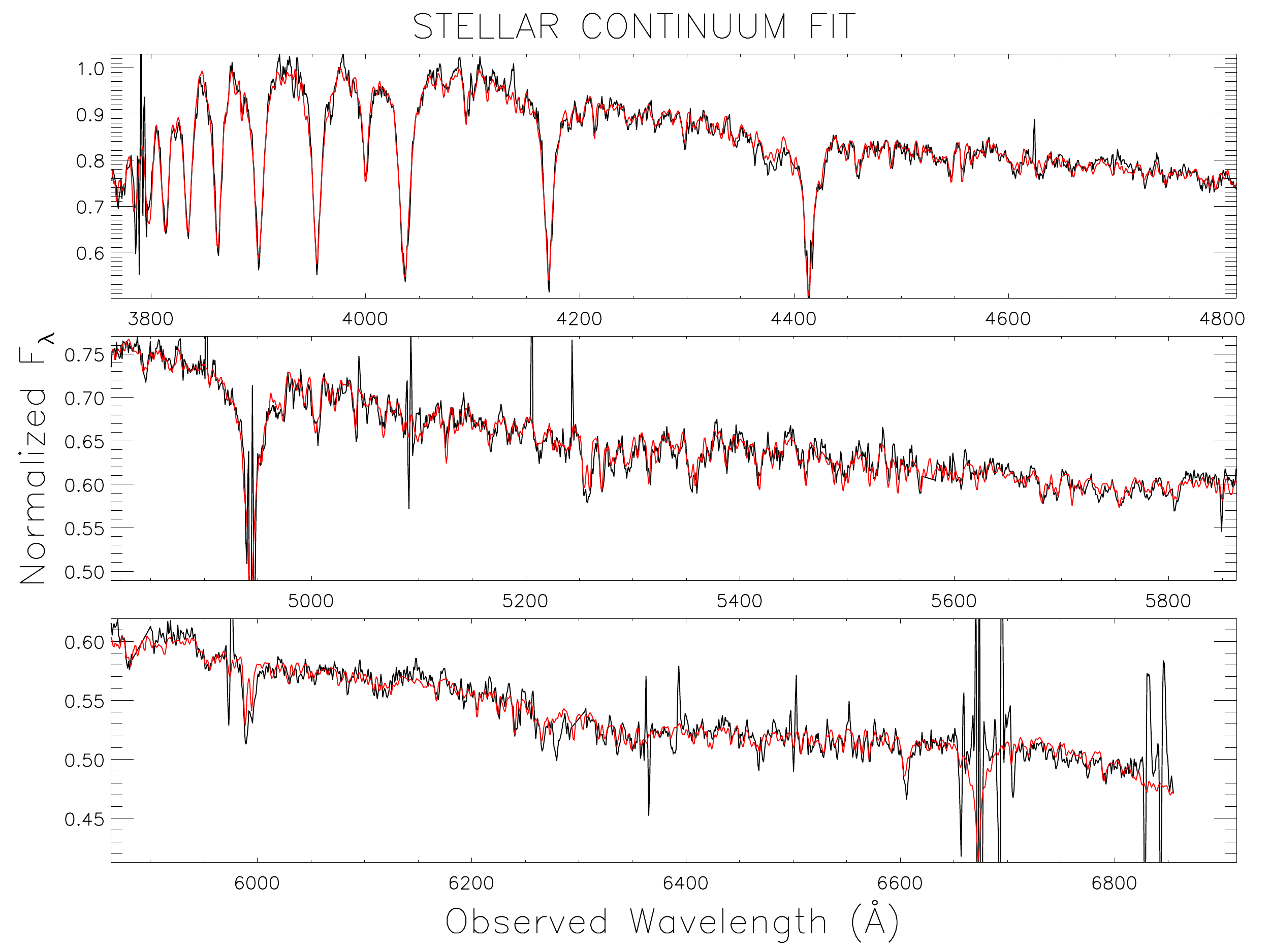,width=0.82\linewidth,clip=} \\
  \end{tabular}
  \caption{\tiny{Example fits to a high-flux and low-flux \htwo\
      region spectrum from our sample (the latter has 1\%\ of the \ha\
      flux of the former).  These demonstrate the high fidelity of our
      fits over a large range of signal-to-noise.  The top panels show
      close-ups of various emission line fits, with the residuals
      underneath each fit.  The bottom panels show the fits to the
      stellar continuum, with the emission-line fits subtracted.
      Black is our data, and red is the fit.  The spectra are shown in
      the observed frame.}}
  \label{fig:example_fit}
\end{figure}
\setcounter{figure}{1}
\begin{figure}
  \centering
  \begin{tabular}{c}
    \epsfig{file=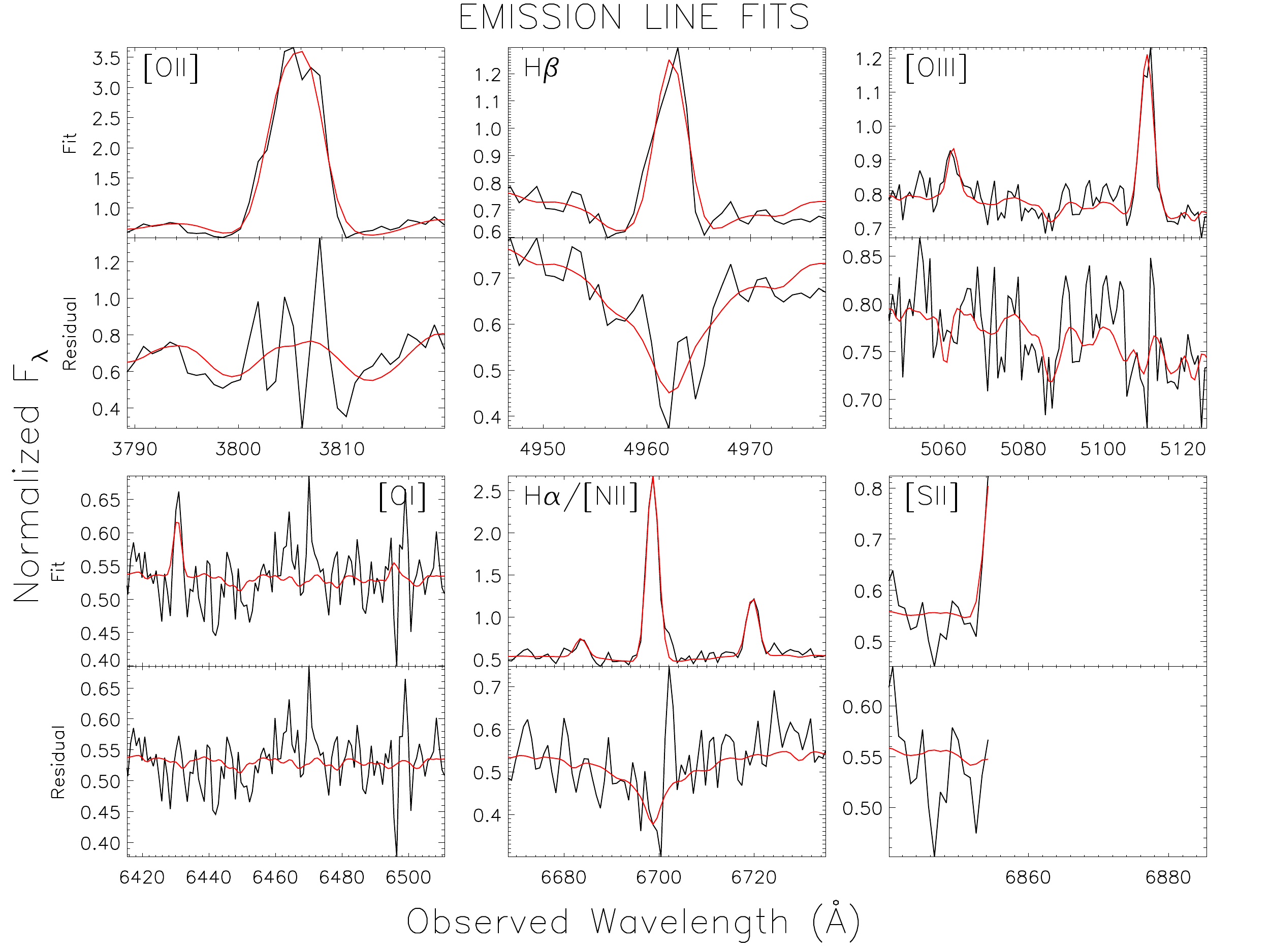,width=0.85\linewidth,clip=} \\
    \epsfig{file=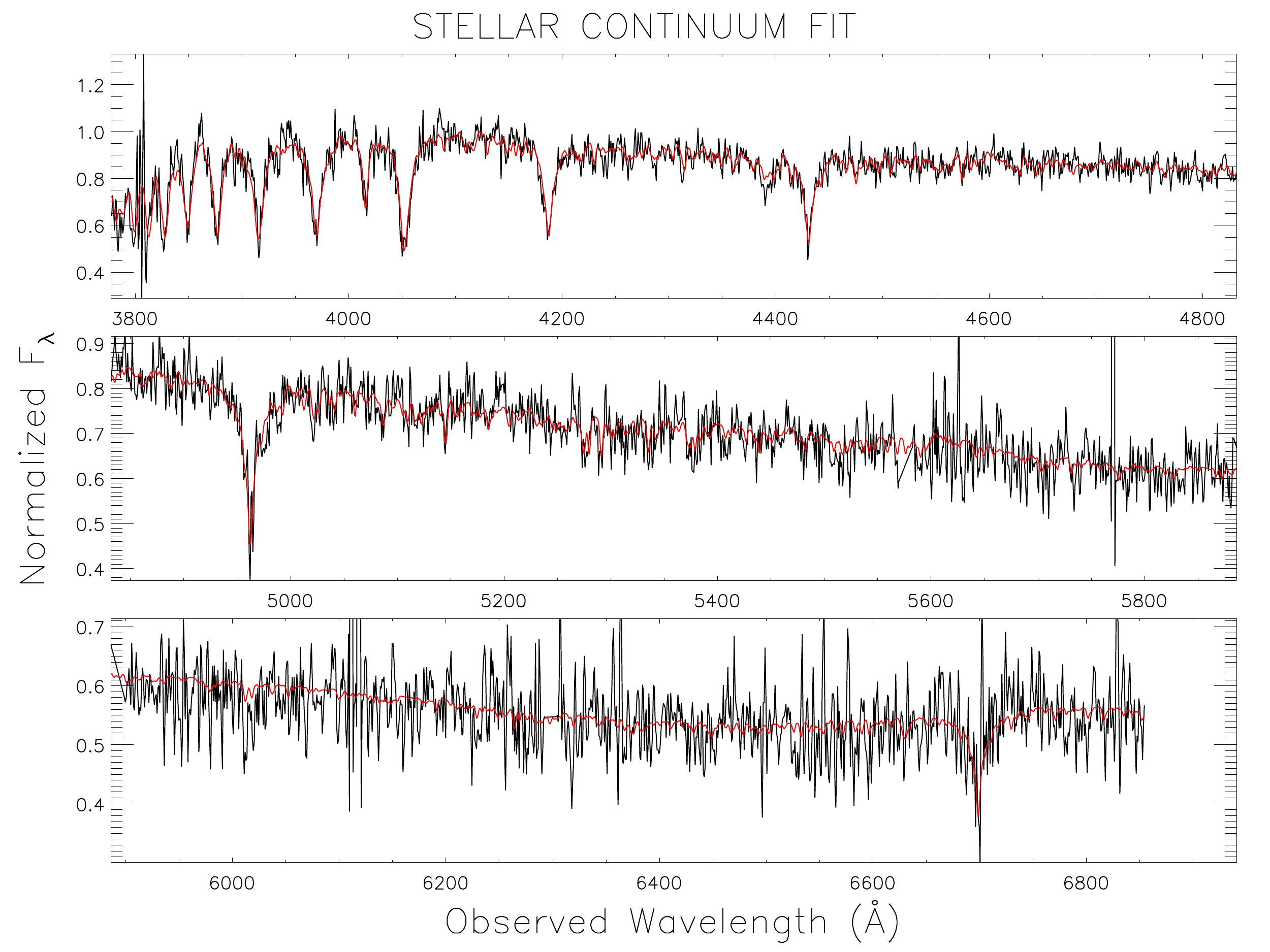,width=0.82\linewidth,clip=} \\
  \end{tabular}
  \caption{\it Continued.}
\end{figure}

\begin{figure}
  \plottwo{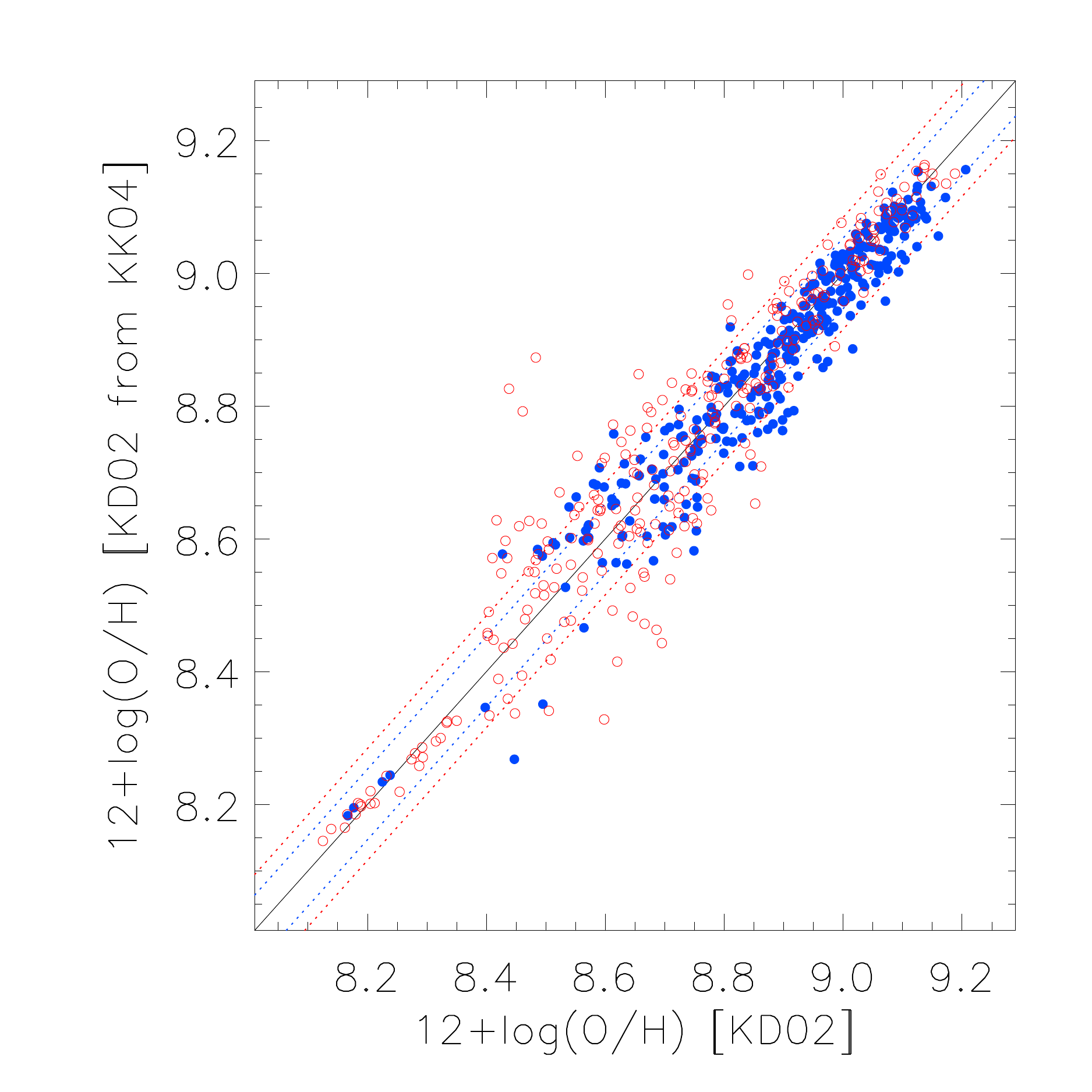}{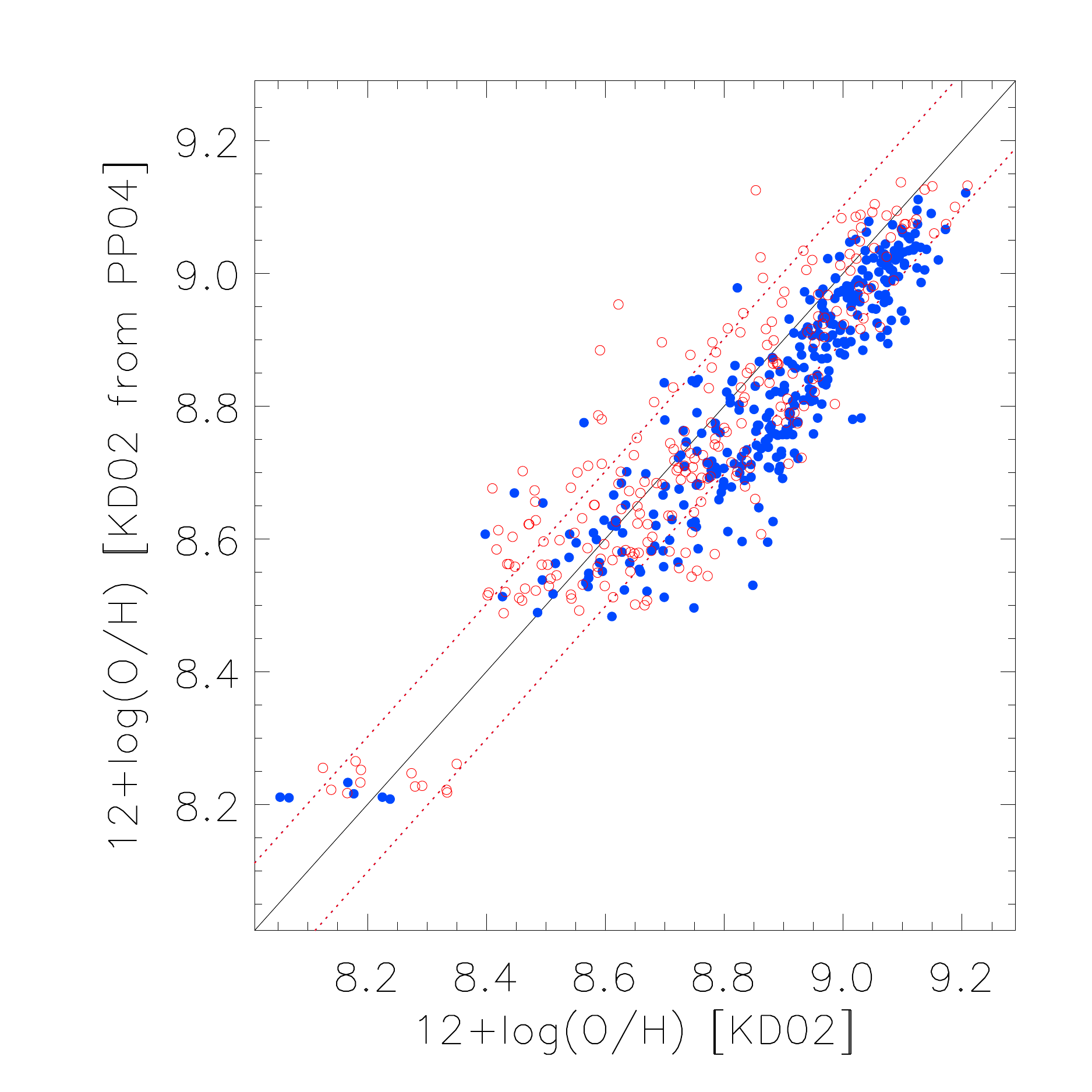}
  \caption{Comparisons between different strong-line methods of
    calculating abundances, within the control sample (red open
    circles) and the interacting sample (blue filled circles).  The
    diagonal black line shows equality, and the blue and red dashed
    lines show the RMS dispersion of each sample ($0.05-0.1$ dex)
    about equality.  We compare our chosen diagnostic (KD02) with KK04
    and PP04, converted into KD02 using \citet{ke08a}
    (\S\ref{sec:spectral-analysis}).  A small shift (0.05~dex) is seen
    toward higher metallicity when comparing KD02 and PP04, but this
    is well within the uncertainties of the calibrations and
    cross-calibrations.  The good agreement between these methods
    points to the absence of strong systematic errors in our
    measurement of line fluxes.}
  \label{fig:abundcomp}
\end{figure}

\clearpage

\begin{figure}
  \plotone{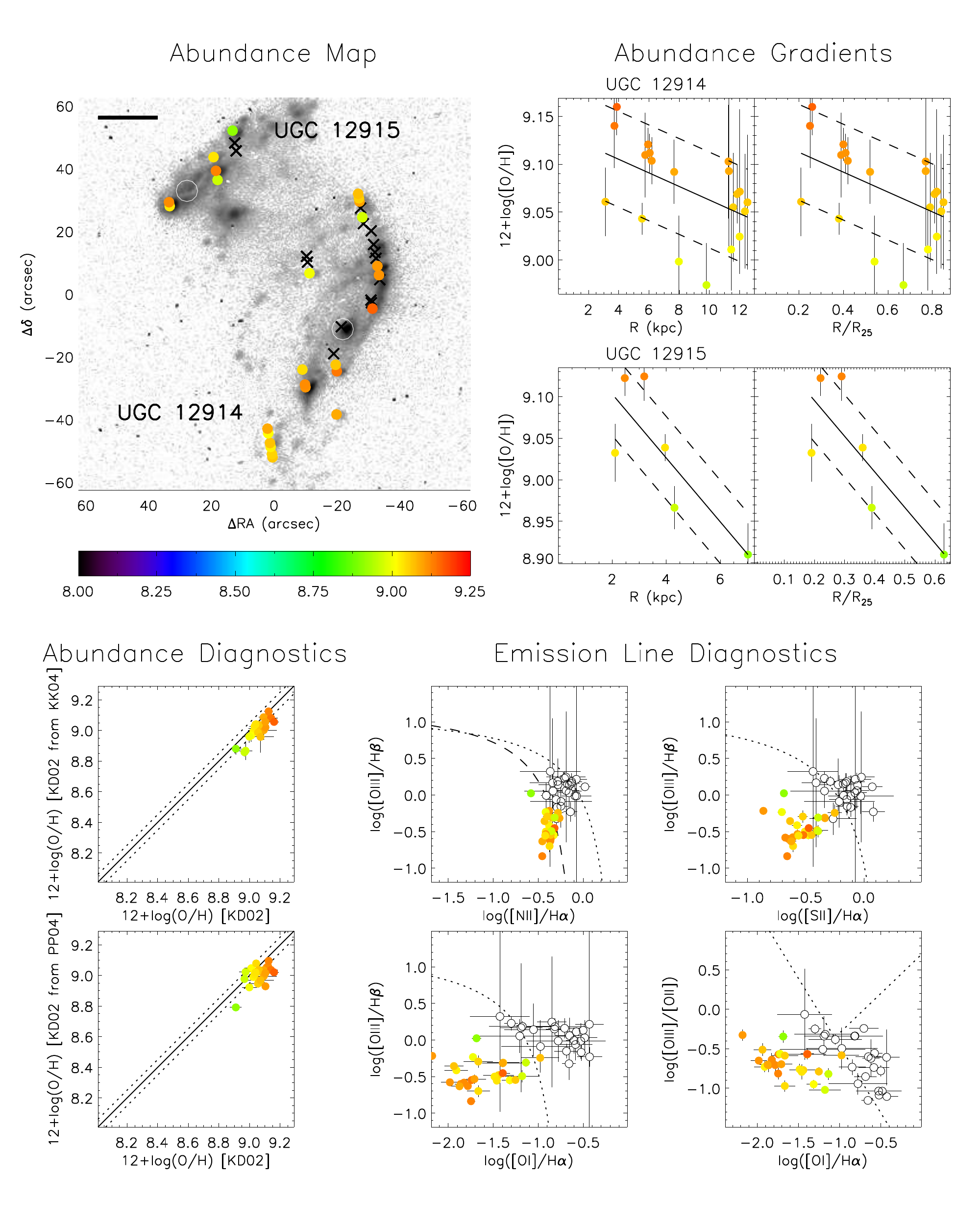}
  \caption{\tiny{Results of emission-line analysis of each system,
      including a map of the oxygen abundance from the \nt/\ot\
      method, overplotted on an \ha\ image (or F435W {\it HST} image
      in the cases of Arp 256 and Arp 298); abundance gradients for
      each galaxy, versus radius in kpc and radius normalized to \rtf;
      comparisons between the KD02 abundance diagnostic and the KK04
      and PP04 diagnostics; and the emission line ratio diagnostics
      from \citet{kewley06a}. In each plot, the data points are
      colored according to the KD02 abundance (see color bar).  The
      images show the near-infrared nuclear position as a white open
      circle, and a horizontal bar in the upper-left corner represents
      5 kpc.  The gradient panels show our formal fit and the average
      RMS (0.05~dex) around this.  For the abundance comparisons, the
      KK04 and PP04 abundances are converted to the KD02 diagnostic
      using the conversions of \citet{ke08a}.  We overplot the line of
      equality and the RMS scatter in the \citet{ke08a} SDSS sample
      after this conversion ($\sim$0.05 dex).  In the emission line
      diagnostics, points used in the abundance calculations are solid
      symbols; points rejected based on the \nt/\ha\ and \oo/\ha\
      diagrams are open symbols.  These rejected points are plotted as
      black crosses on the images.}}
  \label{fig:results}
\end{figure}
\setcounter{figure}{3}
\begin{figure}
  \plotone{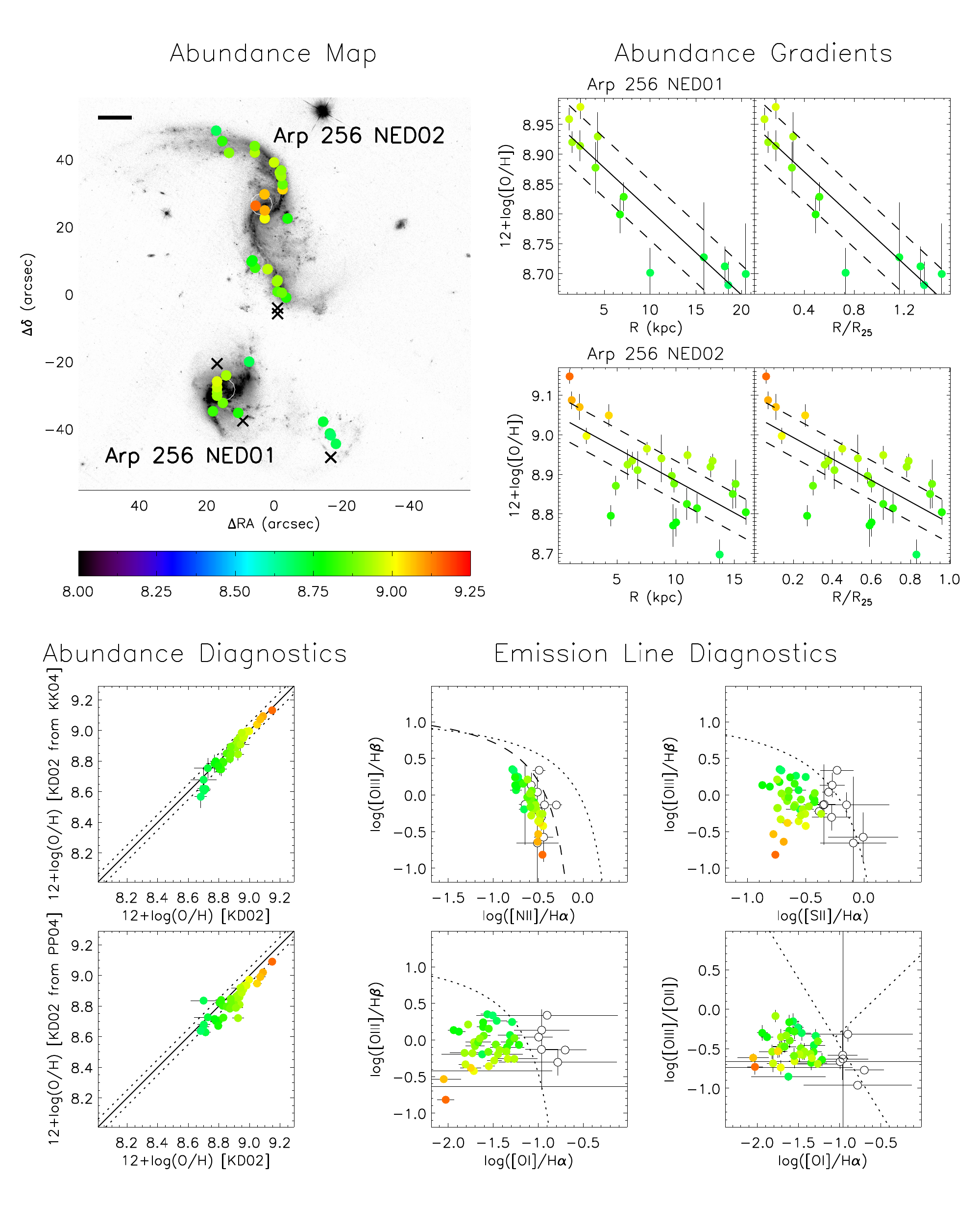}
  \caption{\it Continued.}
\end{figure}
\setcounter{figure}{3}
\begin{figure}
  \plotone{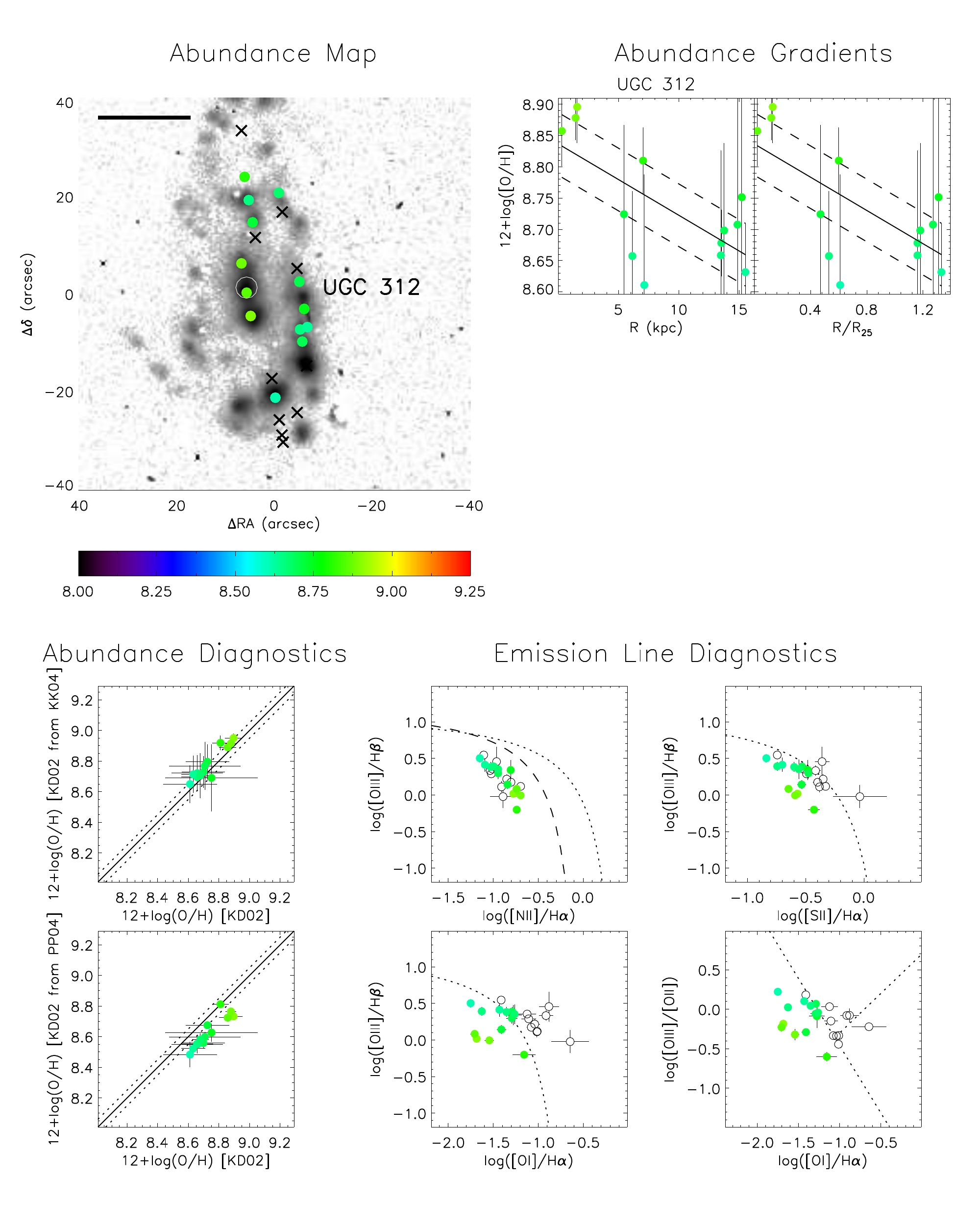}
  \caption{\it Continued.}
\end{figure}
\setcounter{figure}{3}
\begin{figure}
  \plotone{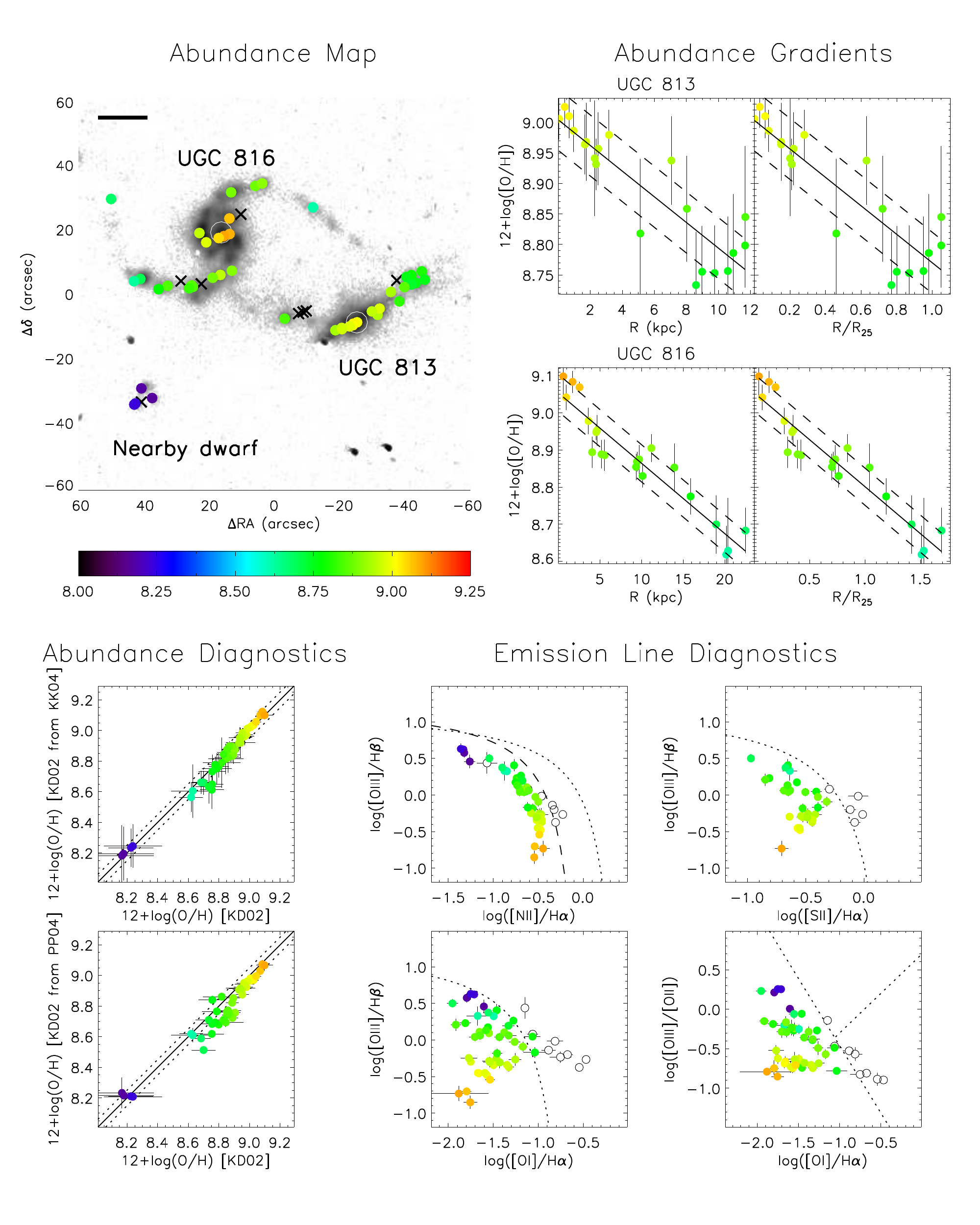}
  \caption{\it Continued.}
\end{figure}
\setcounter{figure}{3}
\begin{figure}
  \plotone{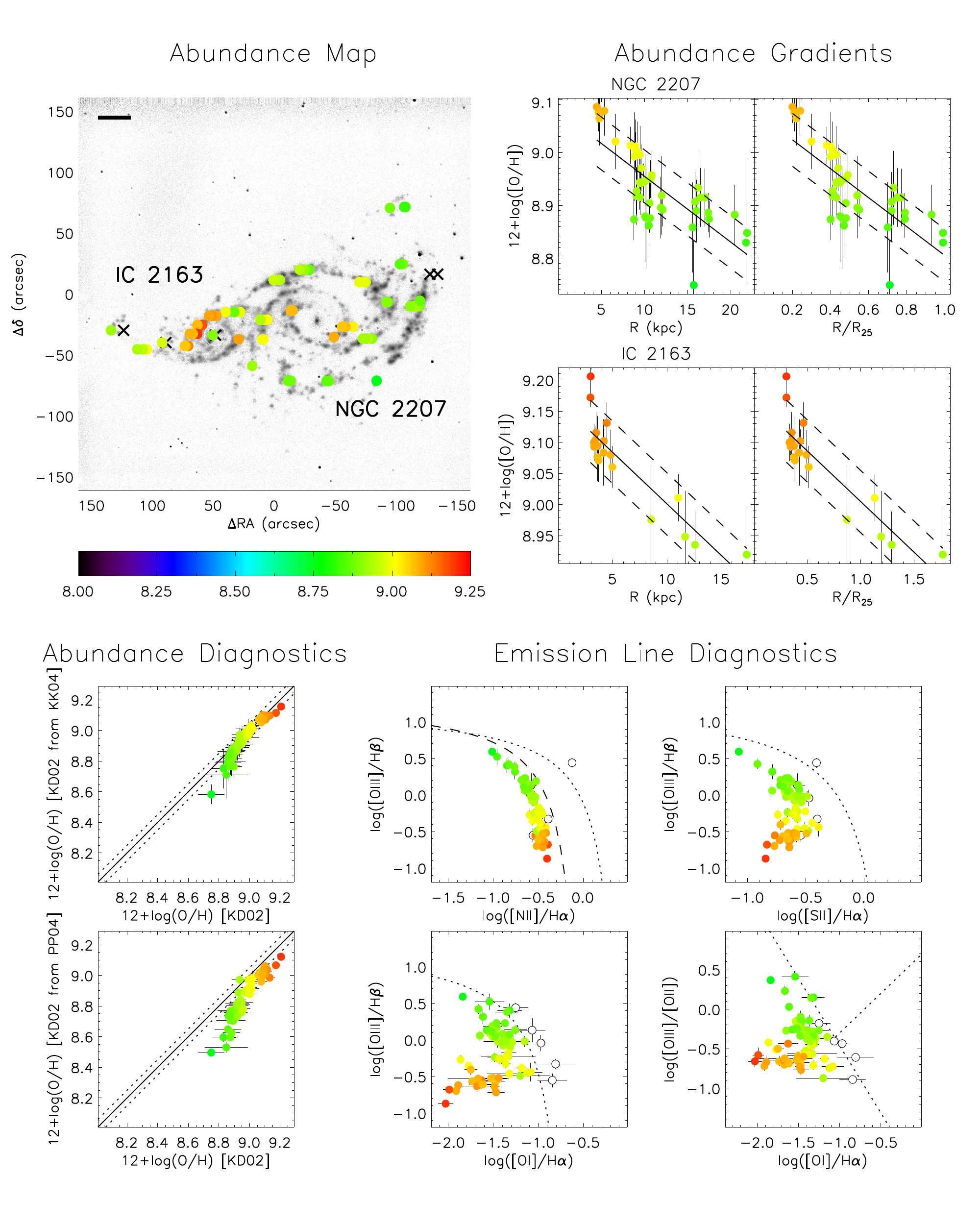}
  \caption{\it Continued.}
\end{figure}
\setcounter{figure}{3}
\begin{figure}
  \plotone{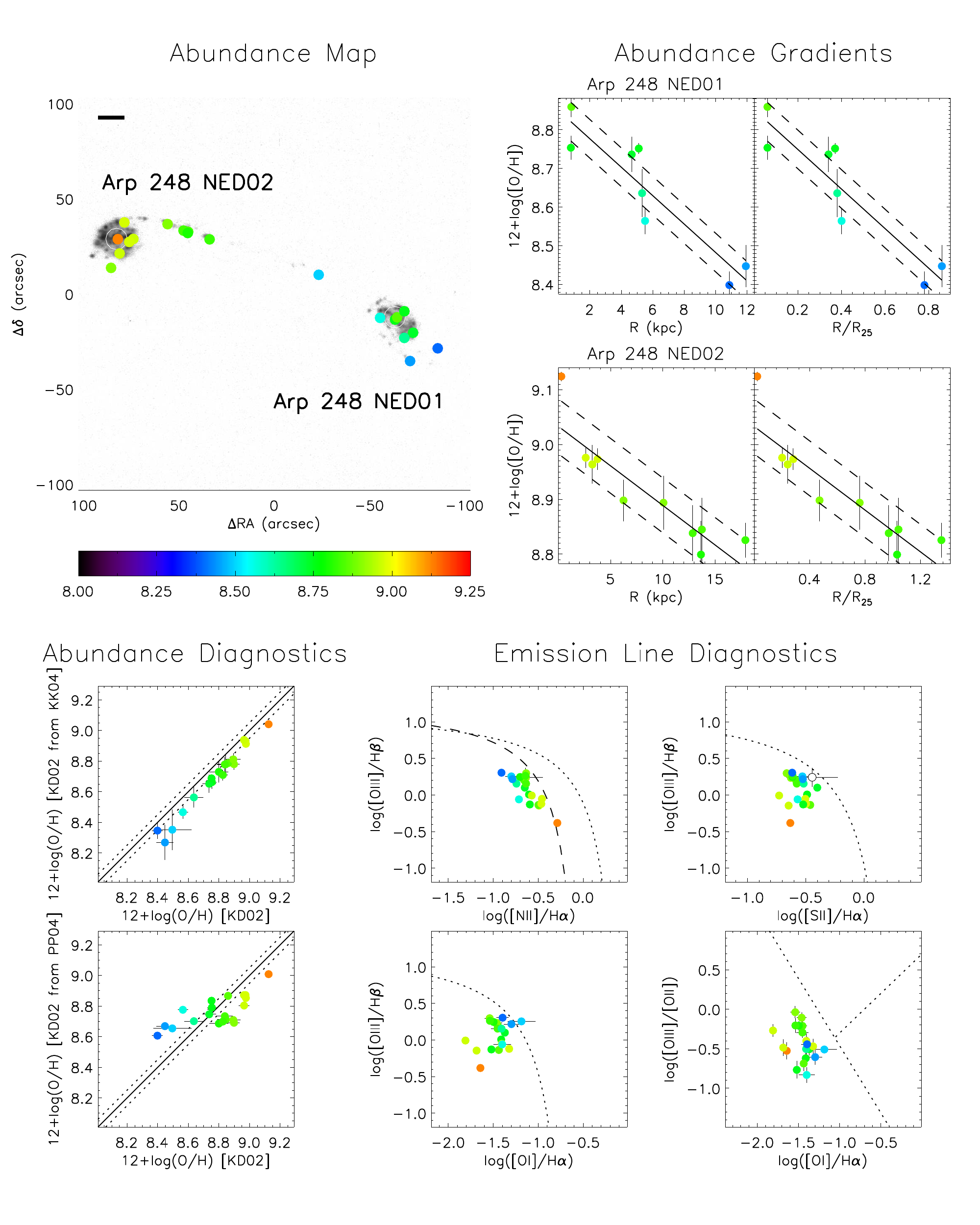}
  \caption{\it Continued.}
\end{figure}
\setcounter{figure}{3}
\begin{figure}
  \plotone{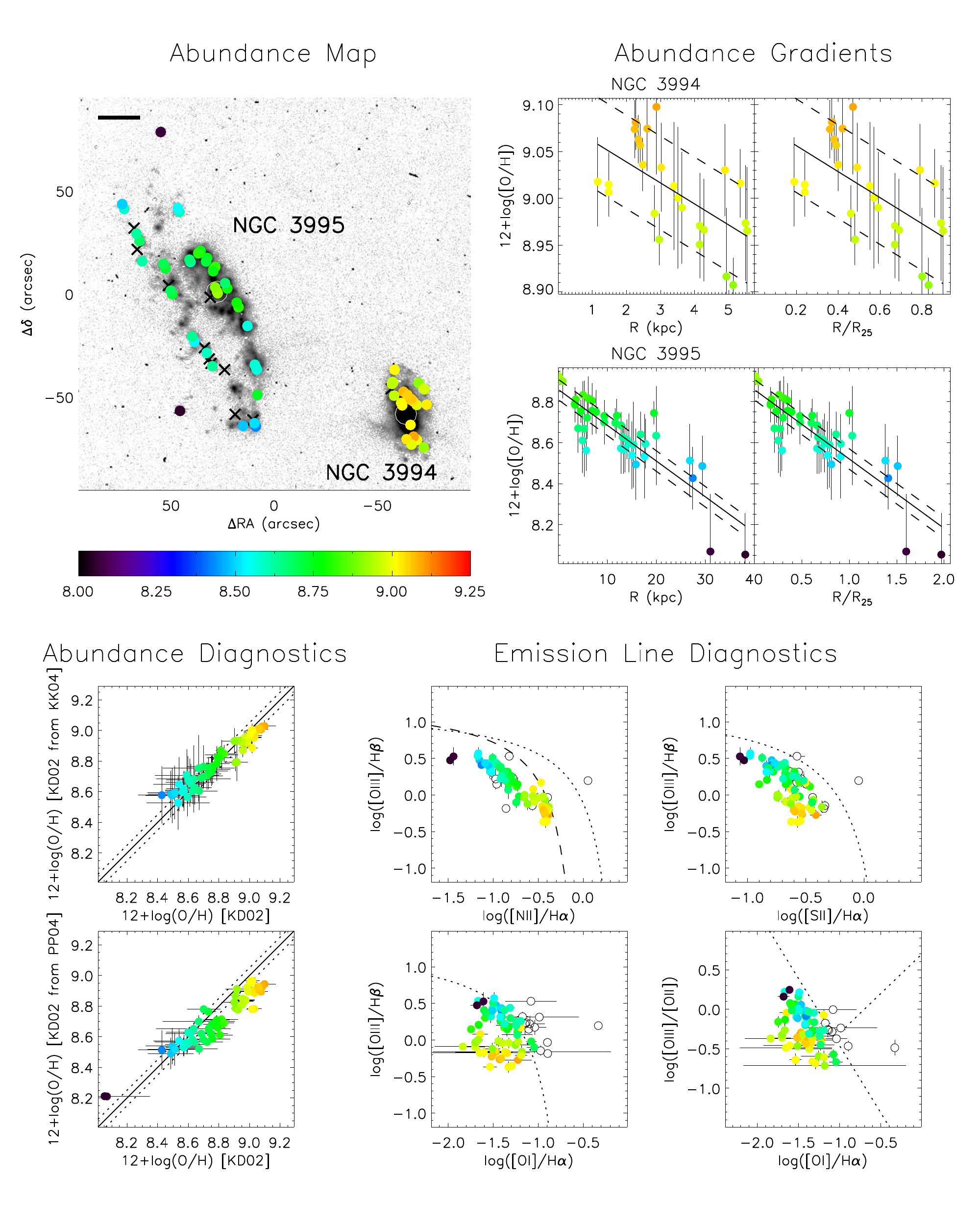}
  \caption{\it Continued.}
\end{figure}
\setcounter{figure}{3}
\begin{figure}
  \plotone{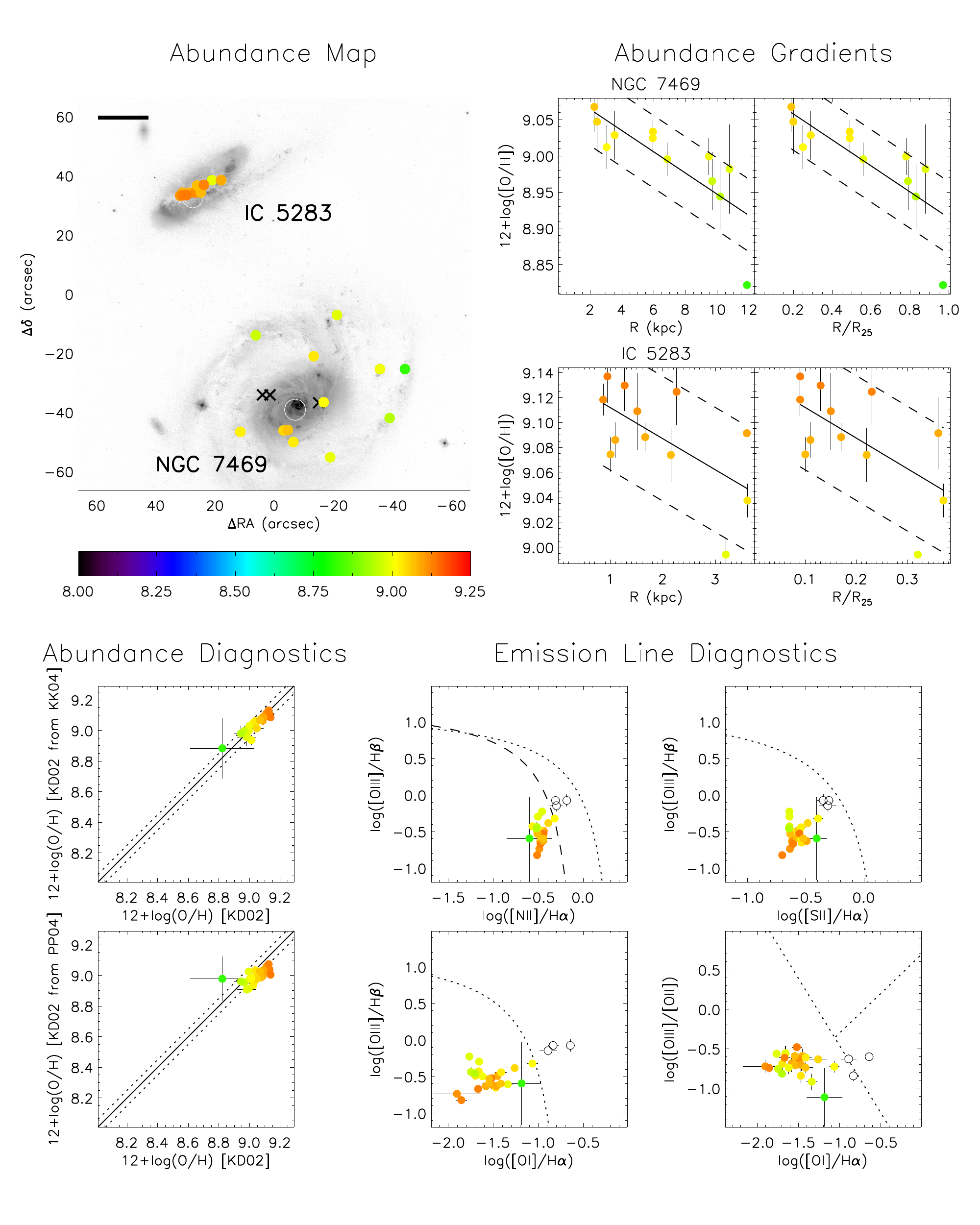}
  \caption{\it Continued.}
\end{figure}
\setcounter{figure}{3}
\begin{figure}
  \plotone{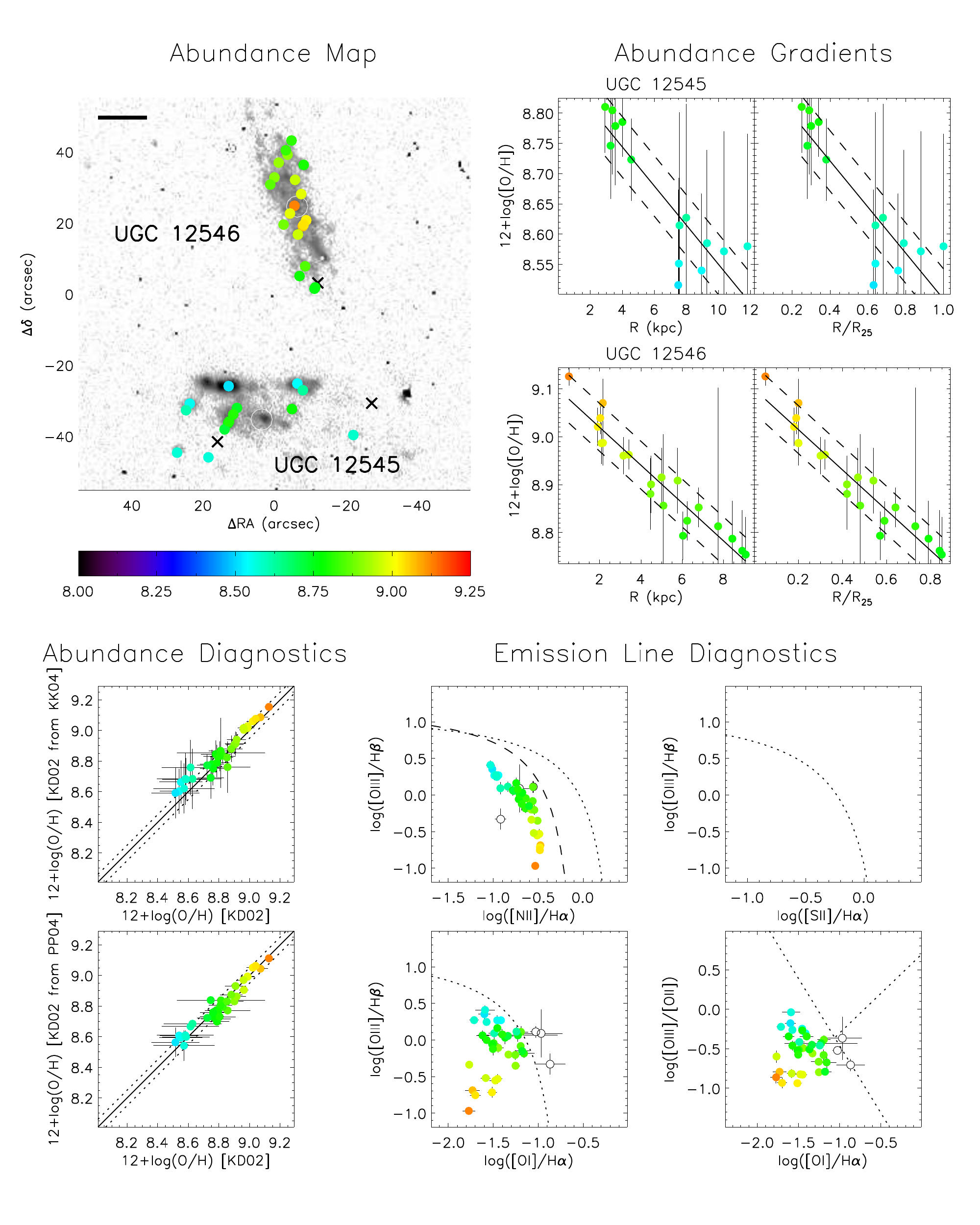}
  \caption{\it Continued.}
\end{figure}

\clearpage

\begin{figure}
  \plottwo{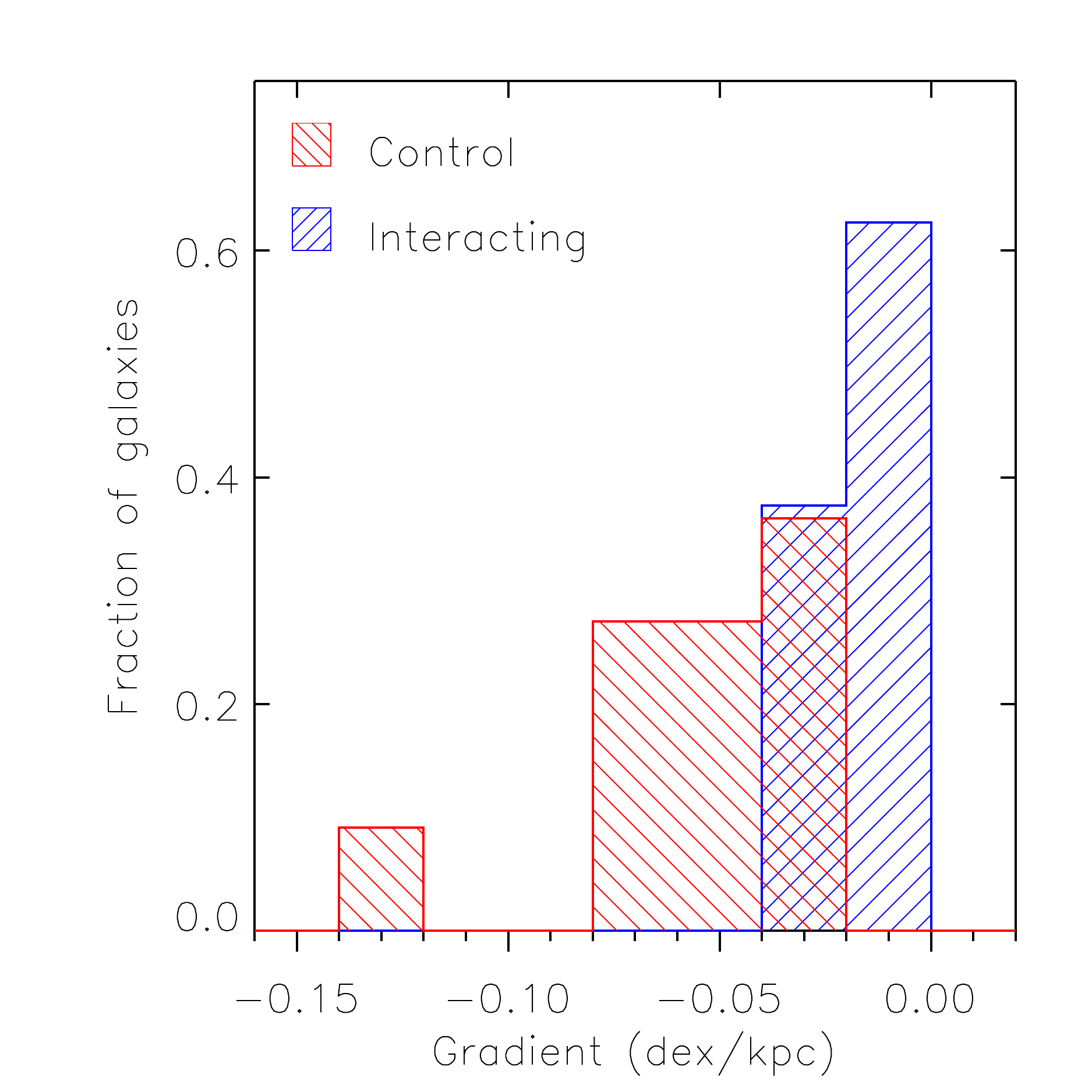}{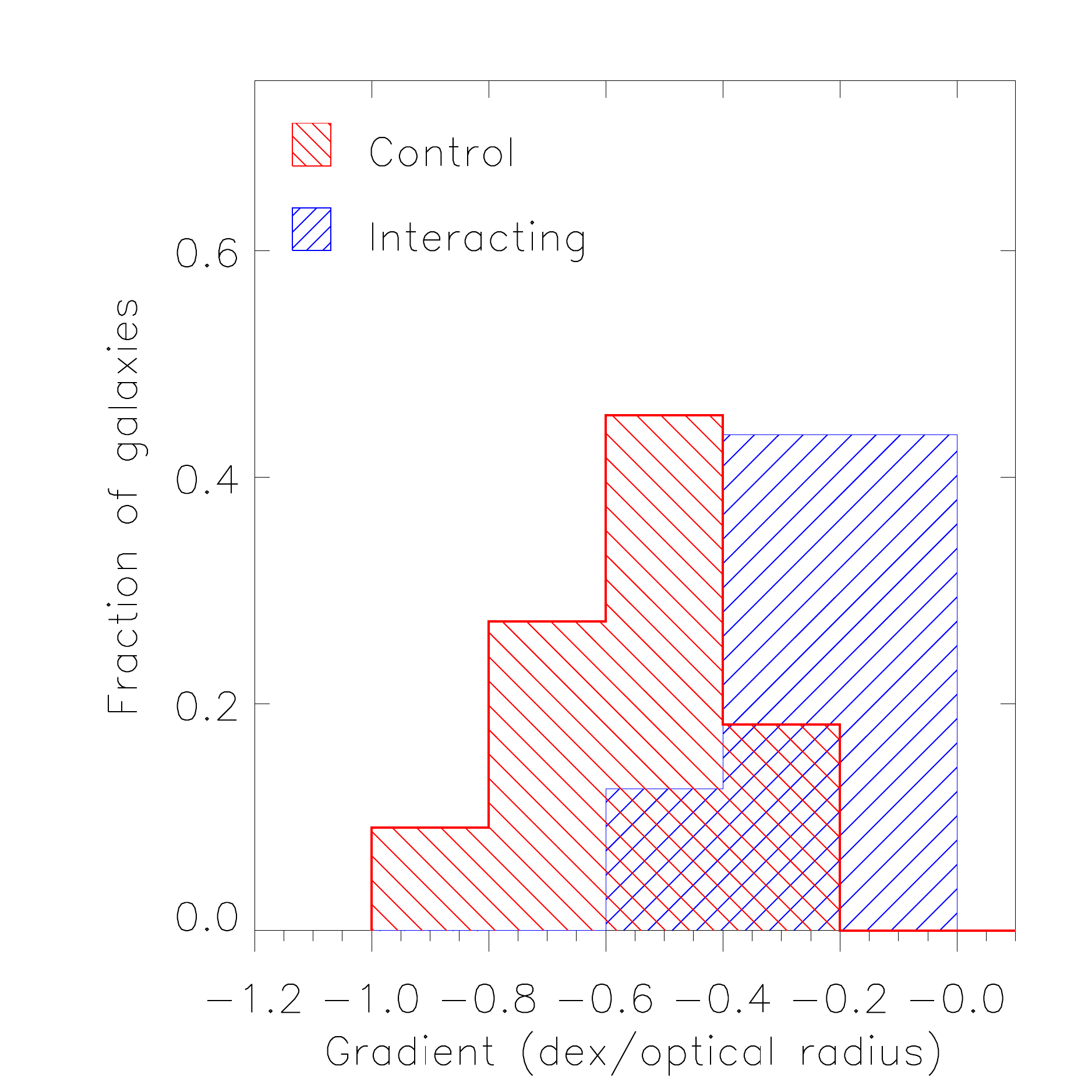}
  \caption{Histograms of gradients in the control and interacting
    samples, in both dex/kpc (left) and dex/\rtf\ (right).  The
    control sample has gradients that are twice as steep as the
    interacting sample.}
  \label{fig:hist-grads}
\end{figure}

\begin{figure}
  \plotone{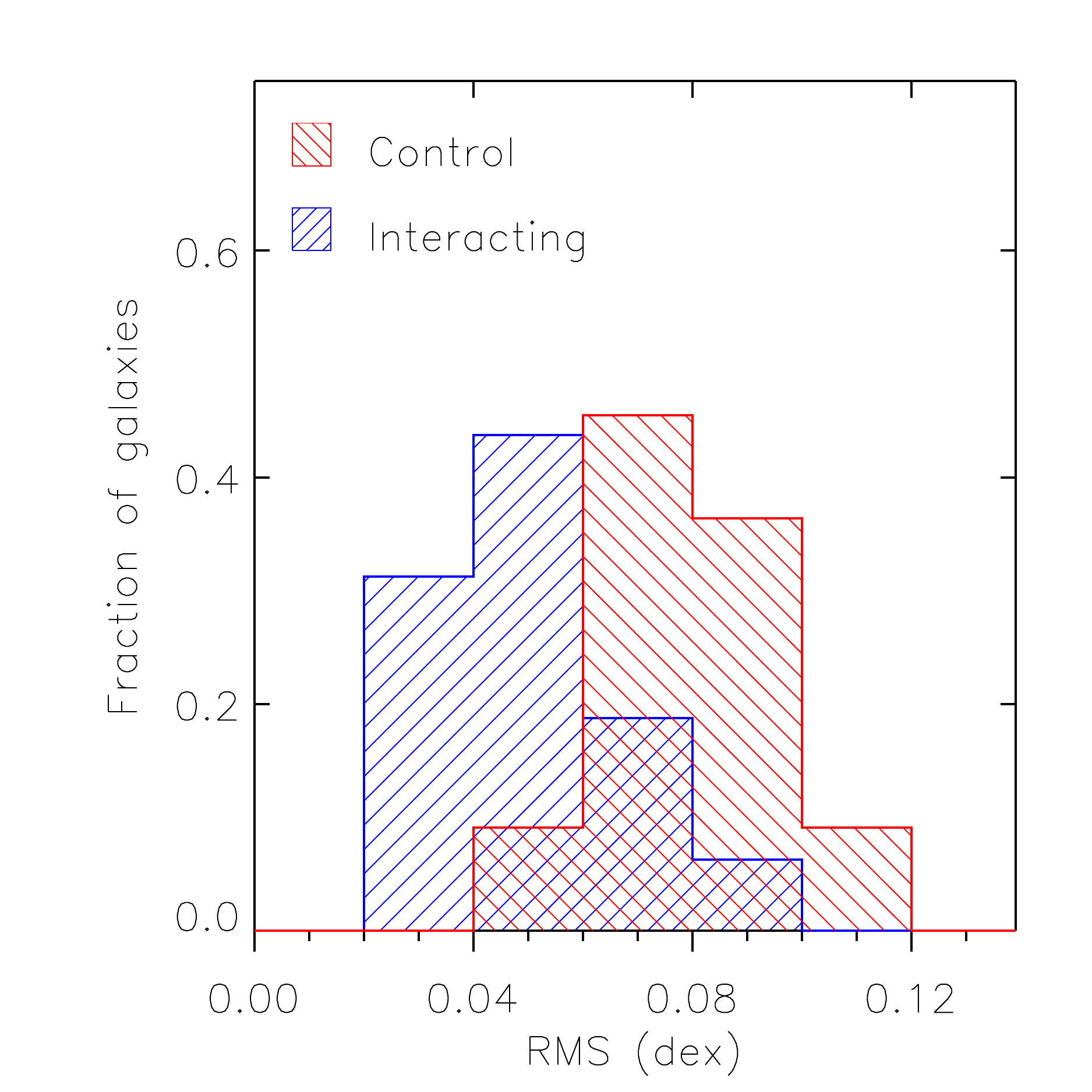}
  \caption{Distribution of RMS residuals from a straight-line
    gradient, for the interacting and control samples.  The
    interacting sample shows smaller, rather than larger, RMS
    residuals.  This implies that the metallicity scatter in galaxies
    does not increase due to a merger; the observed decrease in the
    scatter is likely an artifact of the relative homogeneity of our
    data compared to the control sample.}
  \label{fig:hist-rms}
\end{figure}

\begin{figure}
  \plotone{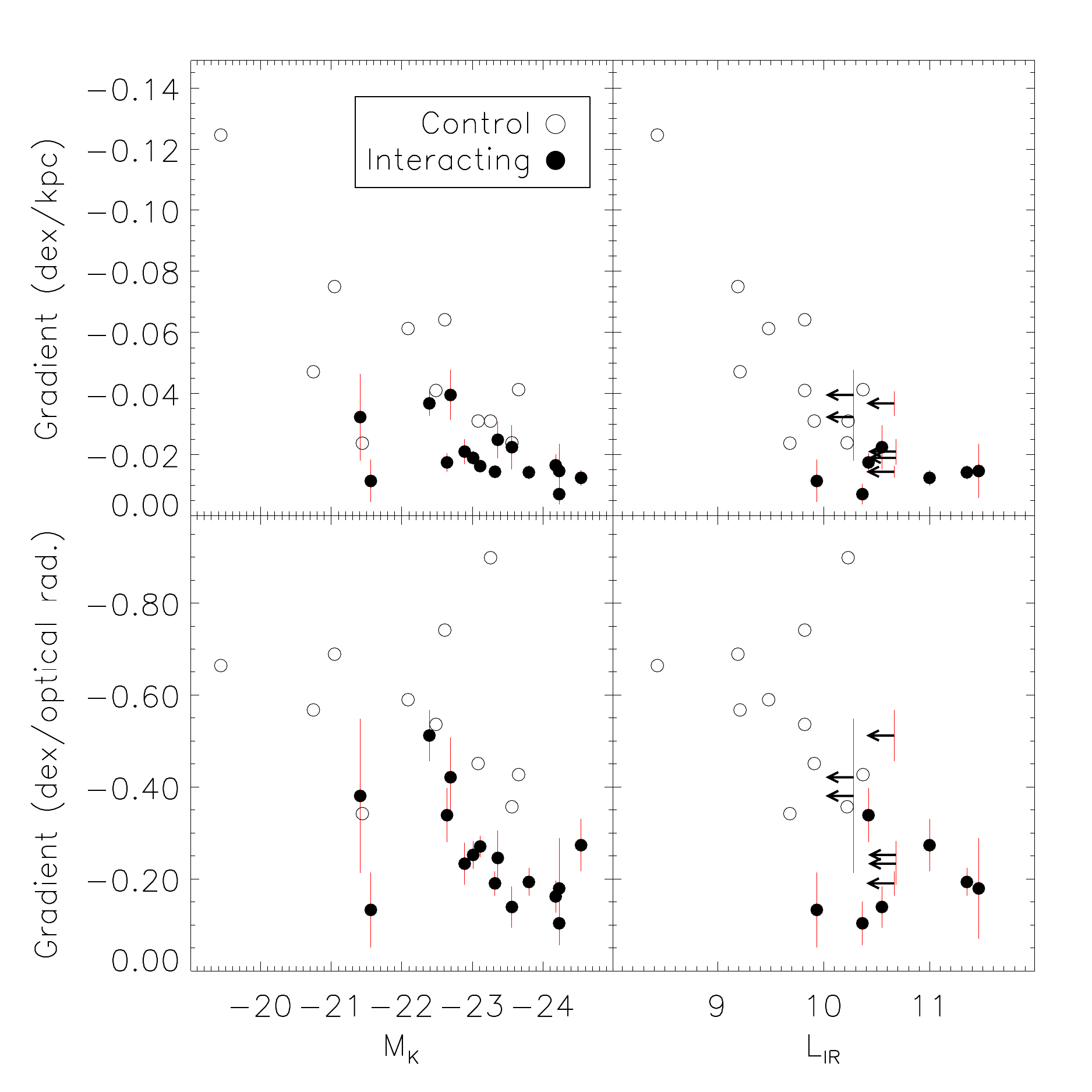}
  \caption{Gradients vs. galaxy properties: $K$-band and total
    infrared luminosities.  Open circles are the control sample, and
    closed circles are the interacting sample.  Arrows represent
    systems with unresolved infrared luminosity; in these cases the
    total system luminosity is an upper limit to the luminosity of
    each galaxy.  No significant trends are apparent within either
    subsample.}
  \label{fig:grads-vs-gal}
\end{figure}

\begin{figure}
  \plotone{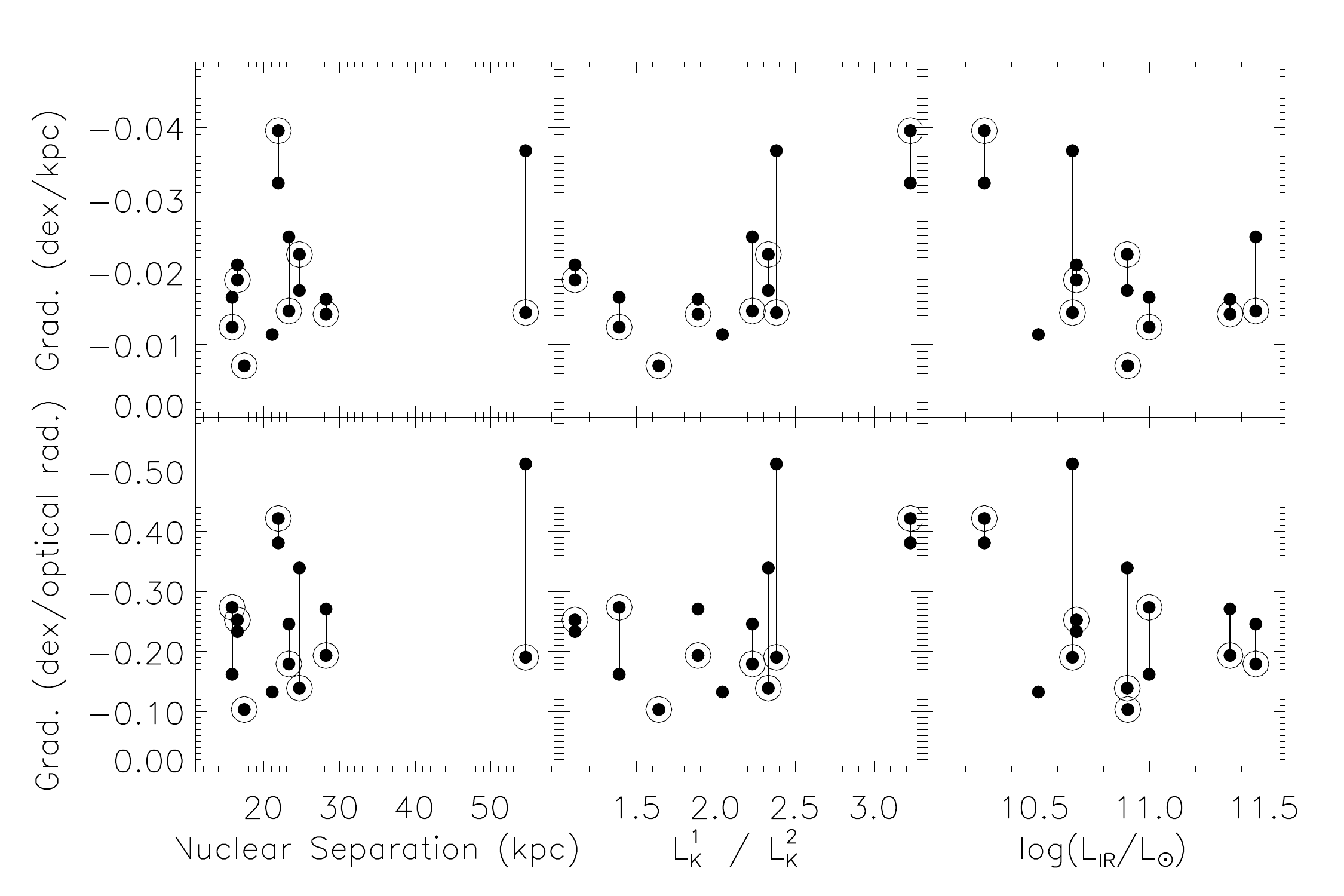}
  \caption{Gradients vs. system properties: nuclear separation, mass
    ratio (as traced by $K$-band luminosity), and total infrared
    luminosity.  Pairs are connected by straight lines, and the most
    massive galaxy in each pair is circled.  No significant trends are
    apparent.}
  \label{fig:grads-vs-sys}
\end{figure}

\begin{figure}
  \plotone{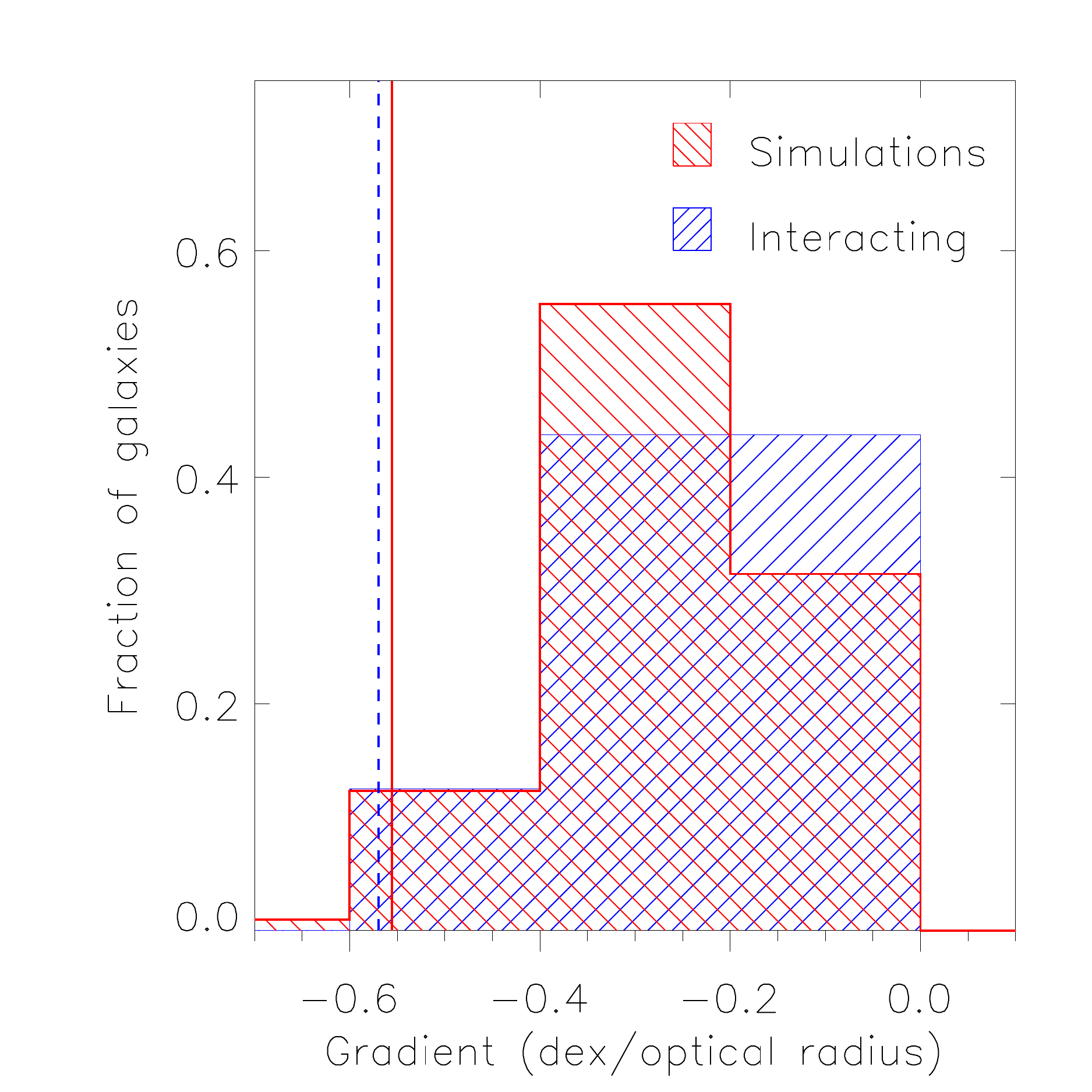}
  \caption{Distribution of gradients in the interacting sample
    compared to the distribution predicted from the simulations of
    \citet{rupke10a}.  The blue, dashed vertical line shows the
    average gradient in the control sample, and the red, solid
    vertical line shows the initial gradient for each simulation.  The
    simulated distribution is computed assuming that the galaxies lie
    between first and second pericenter.  Very good agreement is
    evident.}
  \label{fig:hist-grads-sims}
\end{figure}

\begin{figure}
  \plotone{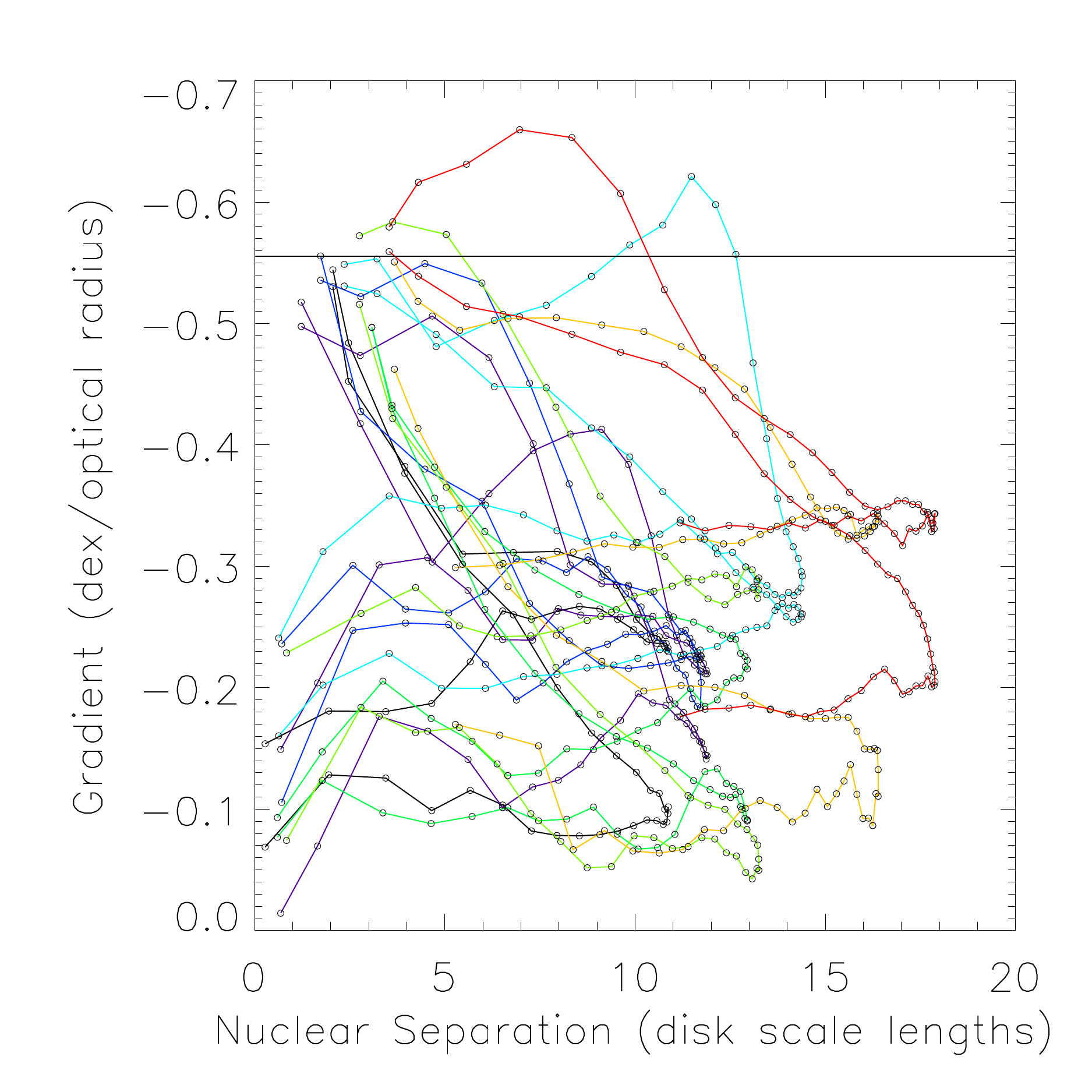}
  \caption{Simulated dependence of gradient on (actual) nuclear
    separation between first and second pericenter \citep{rupke10a}.
    Each point represents a time step for one of two galaxies in one
    of eight simulations.  Points are connected for each galaxy, with
    a different color assigned to each simulation.  The horizontal
    solid line is the initial gradient for each simulation.  At the
    statistical level, no clear dependence of gradient on nuclear
    separation is expected, consistent with what is observed (Figure
    \ref{fig:grads-vs-sys}).  It is also evident that in most systems,
    the gradient flattens quickly after first pericenter, as discussed
    in \citet{rupke10a}.}
  \label{fig:grads-vs-nsep-sims}
\end{figure}

\begin{figure}
  \plotone{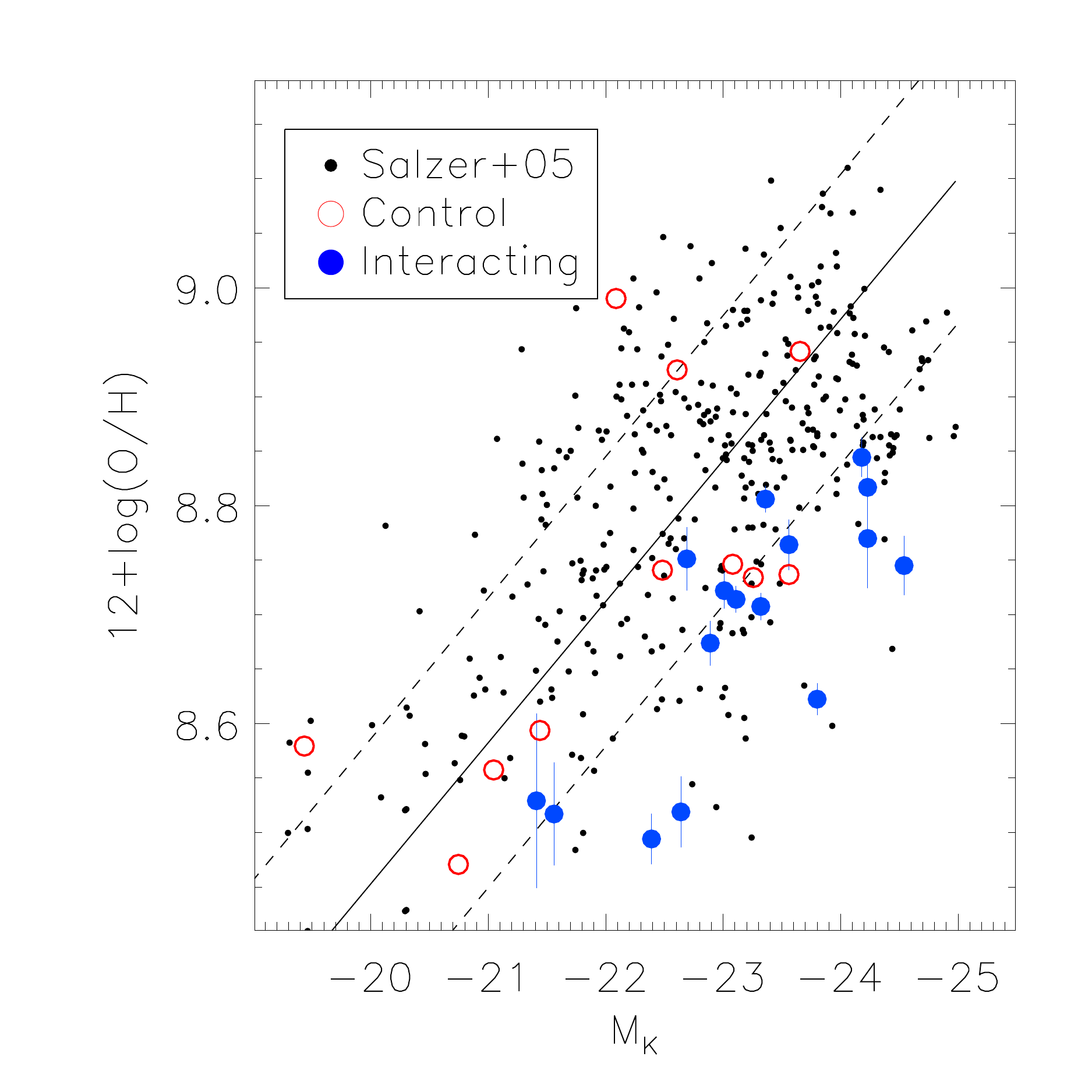}
  \caption{Near-infrared luminosity-metallicity relation.  The small
    black points are emission-line galaxies from \citet{salzer05a};
    the open red circles are the control sample; and the filled blue
    circles are the interacting galaxies.  Nuclear abundances for our
    sample are computed at $R = 0.1$\rtf.  The solid line is the
    bisector of the Y on X and X on Y unweighted least-squares fits to
    the \citet{salzer05a} sample, with RMS dispersion illustrated by
    the dashed lines.  The control and interacting samples are shifted
    down by 0.3~dex so that the control sample lies atop the \lz\
    relation.  The interacting sample falls well beneath the $\lz$
    relation delineated by the control and \citet{salzer05a} samples.}
  \label{fig:lz-0p1}
\end{figure}

\begin{figure}
  \plotone{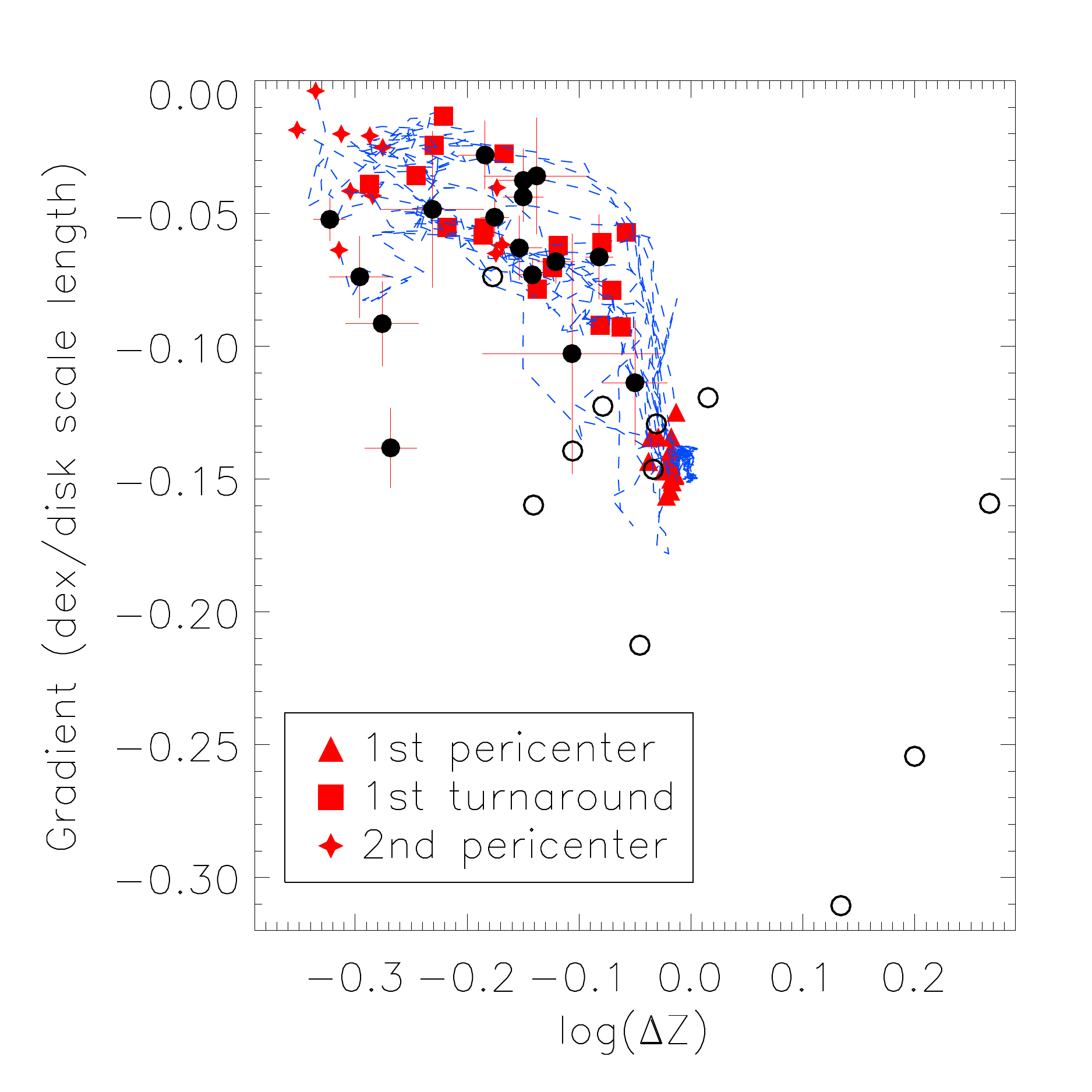}
  \caption{Estimated offset from the \lz\ relation vs. metallicity
    gradient.  Open circles are the control sample, and filled circles
    are the interacting sample.  Dashed blue lines illustrate the time
    evolution in this phase space of the eight major merger models
    from \citet{rupke10a}.  The models move from the center to the
    upper left.  The triangles, squares, and star indicate various
    important points in the evolution of the simulated merger.  The
    input gradient for the simulations is chosen to match the average
    gradient per stellar disk scale length from the control sample.
    Given both the measurement and systematic uncertainties in this
    procedure, there is excellent agreement between the data and
    models.  The data are consistent with the typical system lying
    near first turnaround.}
  \label{fig:dz-vs-grad}
\end{figure}

\begin{figure}
  \plotone{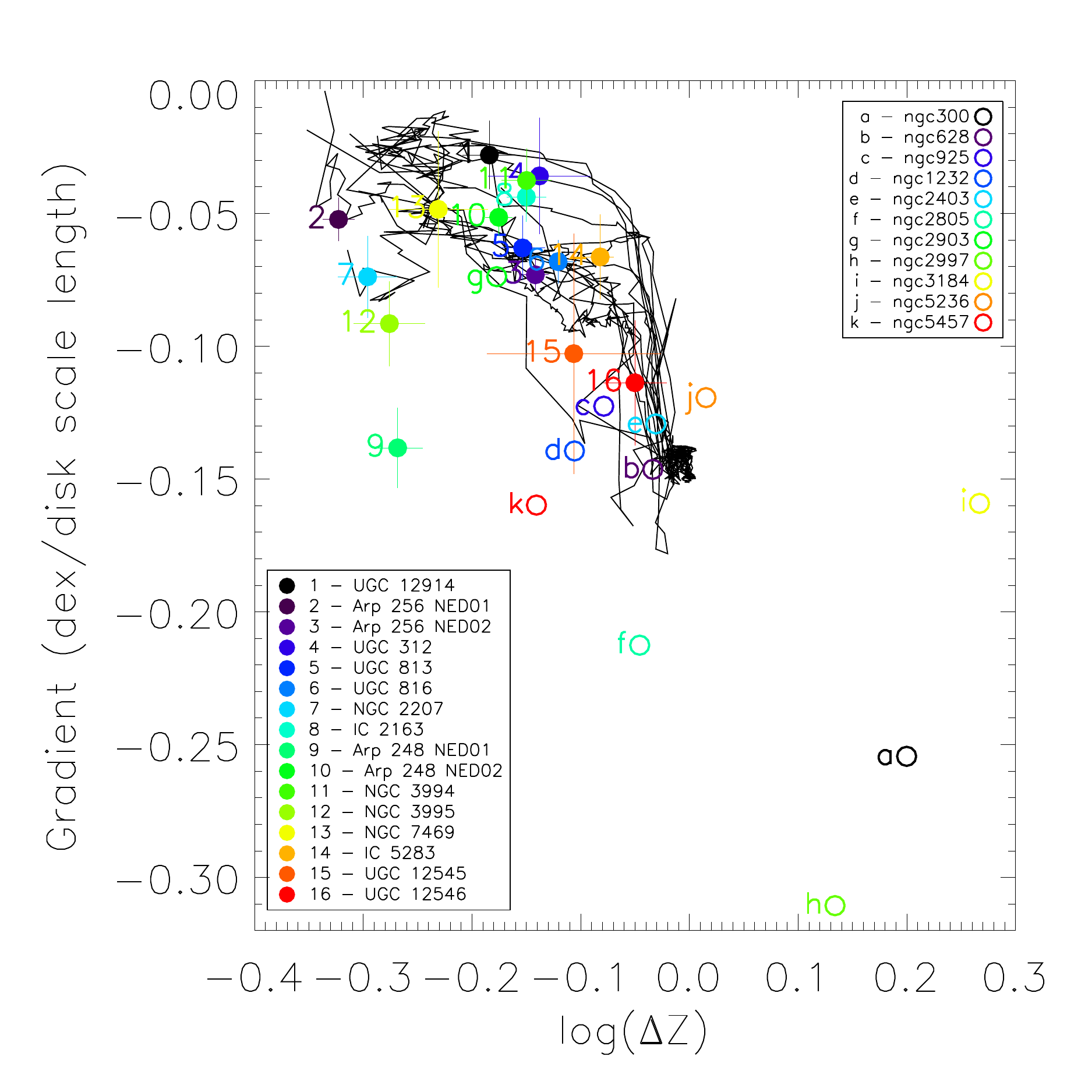}
  \caption{Estimated offset from the \lz\ relation vs. metallicity
    gradient.  See Figure \ref{fig:dz-vs-grad} for more details.
    Individual galaxies are indicated.}
  \label{fig:dz-vs-grad-lab}
\end{figure}

\begin{figure}
  \plotone{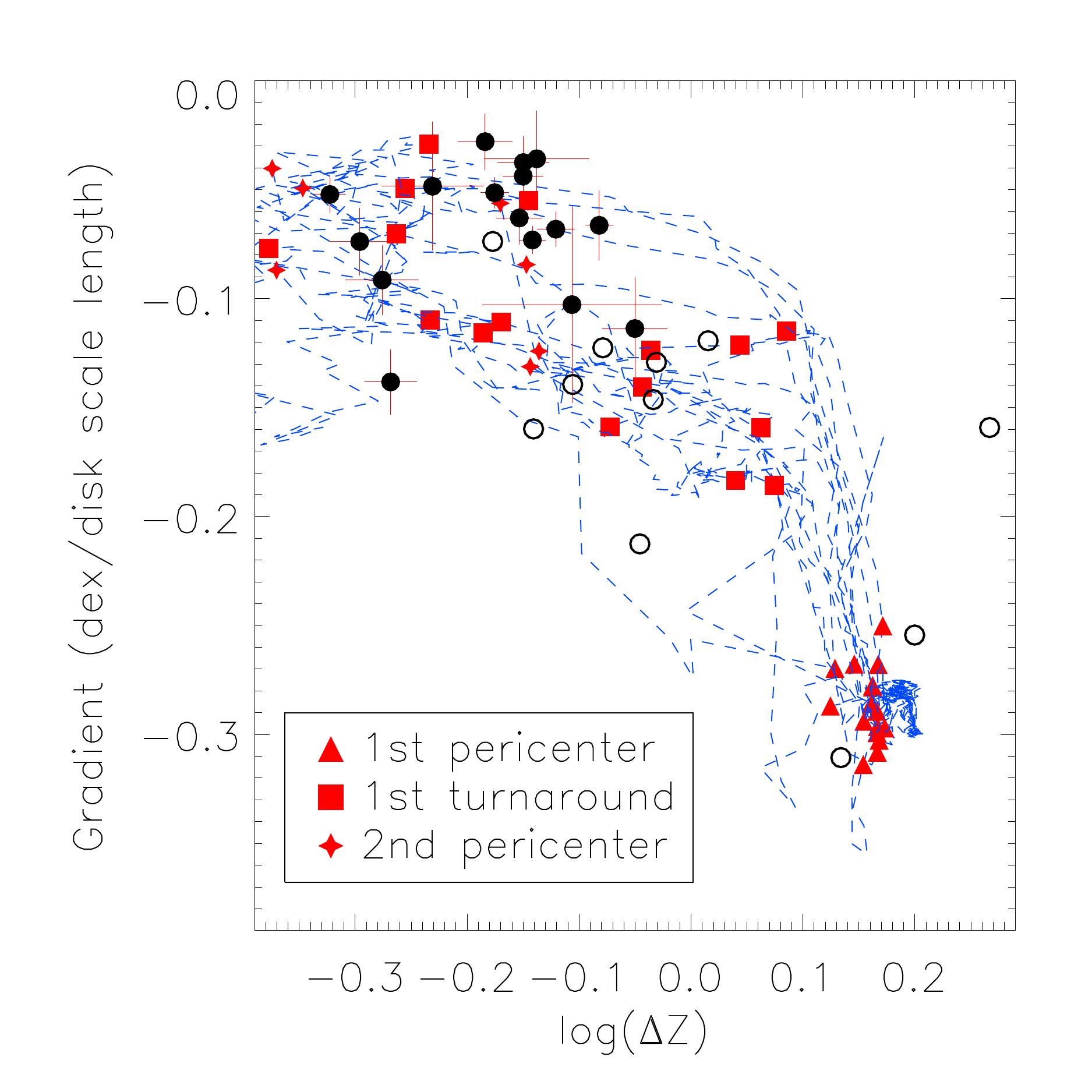}
  \caption{Estimated offset from the \lz\ relation vs. metallicity
    gradient.  See Figure \ref{fig:dz-vs-grad} for more details.  In
    the simulations displayed here, we used for initial conditions the
    steepest gradients observed in the control sample, and the
    corresponding \lz\ relation offsets.  The overlap between data and
    simulations is poorer than in Fig. \ref{fig:dz-vs-grad}, and this
    scenario would require most of the pairs to lie near second
    pericenter rather than first turnaround.  The data thus favor the
    shallower progenitor gradients illustrated in
    Fig. \ref{fig:dz-vs-grad}.}
  \label{fig:dz-vs-grad-steep}
\end{figure}

\begin{figure}
  \plottwo{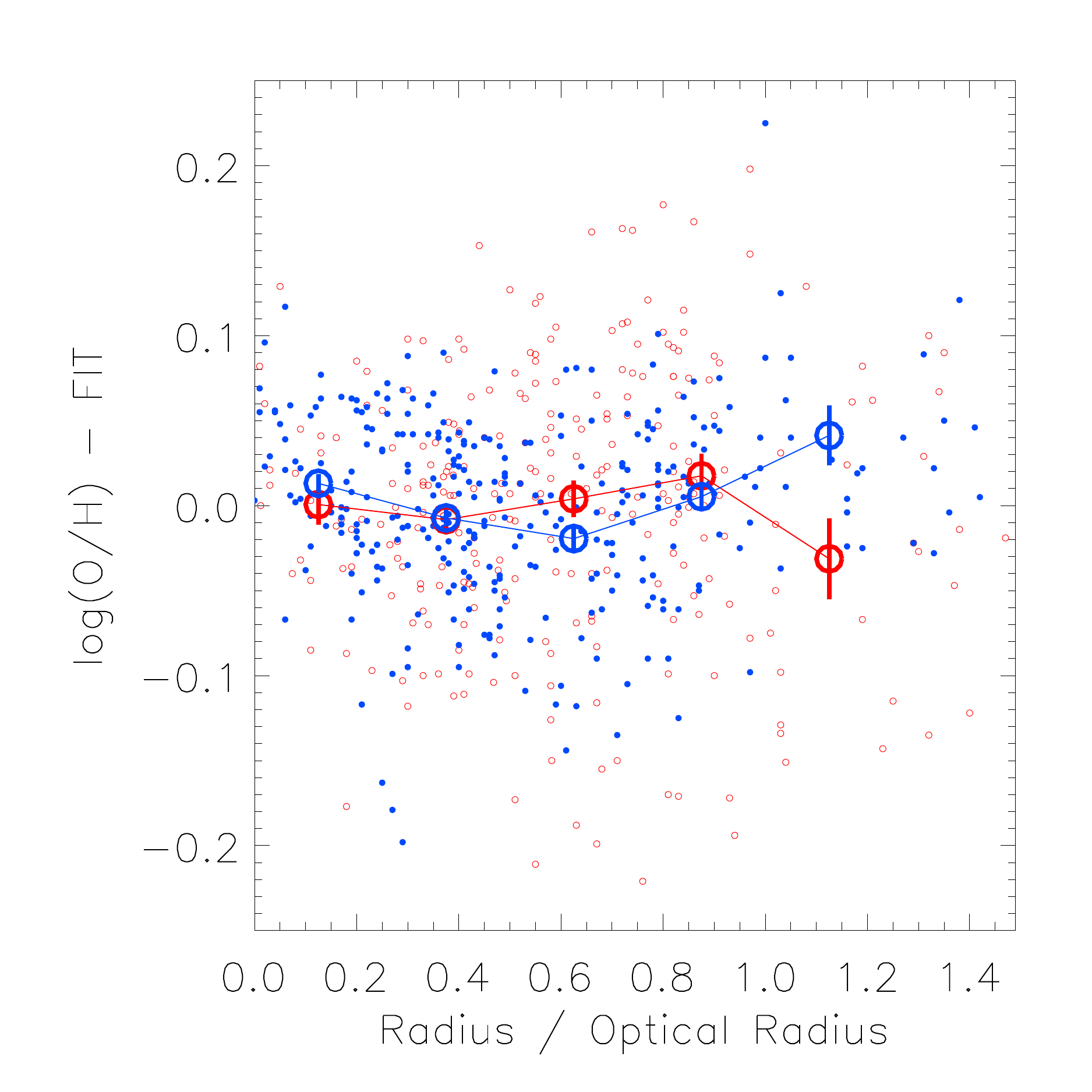}{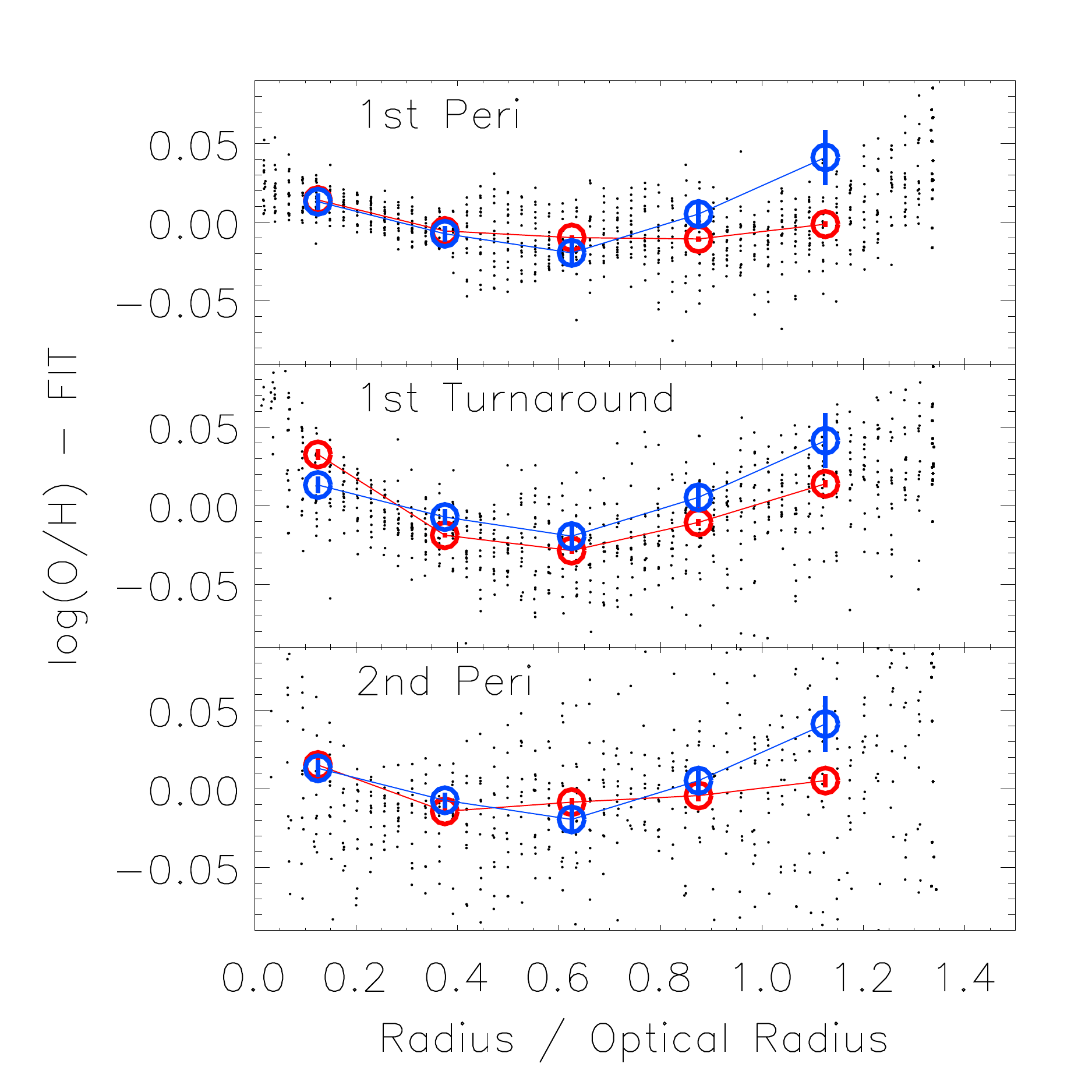}
  \caption{(left) Metallicity residual from a straight line fit
    vs. radius for all \htwo\ regions in the control (open red
    circles) and the interacting (filled blue circles) samples.
    Binned points are averages over bins of size 0.25\rtf, with the
    standard error shown as vertical error bars.  The control sample
    is consistent with a mostly flat profile, while the interacting
    sample shows a ``concave up'' shape.  (right) The same, but for
    the simulations (black points and red, open circles), with the
    binned data for the interacting sample overplotted.  The best
    agreement in shape is found between the interacting sample and the
    simulations at first turnaround, which both have a ``concave up''
    profile residual.  Exact quantitative agreement is lacking, but
    this comparison is consistent with the idea that our sample lies
    on average near first turnaround.}
  \label{fig:profile}
\end{figure}

\clearpage

\begin{deluxetable}{llcccrrrrc}
  \tablecaption{Sample\label{tab:sample}}
  \tablewidth{0pt}

  \tablehead{
    \colhead{Galaxy} & \colhead{System} & \colhead{$cz$} &
    \colhead{Nuc. Sep.} & \colhead{$M_K$} & 
    \colhead{log(\lir/\lsun)} & \colhead{\rtf} &
    \colhead{$i$} & \colhead{PA$_{nodes}$} & \colhead{Ref.} \\
    \colhead{(1)} & \colhead{(2)} & \colhead{(3)} & \colhead{(4)} &
    \colhead{(5)} & \colhead{(6)} & \colhead{(7)} & \colhead{(8)} &
    \colhead{(9)} & \colhead{(10)}
  }

  \startdata
                     UGC 12914 &            VV254 &  4371 & \nodata &   -24.23 &                  10.36 &   14.69 &      54 &     160 &      2,3,4 \\
                     UGC 12915 &            VV254 &  4336 &   17.40 &   -23.70 &                  10.76 &   11.11 &      73 &     135 &      2,3,4 \\
                 Arp 256 NED02 &          Arp 256 &  8193 & \nodata &   -23.11 &                \nodata &   16.58 &  0$-$45 &      70 &      1,3,4 \\
                 Arp 256 NED01 &          Arp 256 &  8125 &   28.20 &   -23.80 &                  11.35 &   13.66 &      61 &      80 &      1,3,4 \\
                       UGC 312 &            HCG 2 &  4364 & \nodata &   -21.56 &                   9.93 &   11.67 &      79 &       8 &          2 \\
      Mrk 552\tablenotemark{a} &            HCG 2 &  4349 &   21.10 &   -22.33 &                  10.38 & \nodata & \nodata & \nodata &          2 \\
      UGC 314\tablenotemark{a} &            HCG 2 &  4271 &   65.40 &   -20.52 &                \nodata & \nodata & \nodata & \nodata &    \nodata \\
                       UGC 813 &          WBL 036 &  5344 & \nodata &   -22.89 & 10.68\tablenotemark{b} &   11.16 &      72 &     110 &          2 \\
                       UGC 816 &          WBL 036 &  5188 &   16.50 &   -23.01 & 10.68\tablenotemark{b} &   13.33 &      62 &     190 &          2 \\
 CGCG 551-011\tablenotemark{a} &          WBL 036 &  5373 &   42.60 &   -23.27 &                \nodata & \nodata & \nodata & \nodata &          2 \\
                      NGC 2207 & NGC 2207/IC 2163 &  2741 & \nodata &   -24.54 &                  11.00 &   22.11 &      35 &     140 &        3,4 \\
                       IC 2163 & NGC 2207/IC 2163 &  2765 &   15.80 &   -24.18 &                \nodata &    9.80 &      35 &     128 &        3,4 \\
                 Arp 248 NED01 &          Arp 248 &  5167 & \nodata &   -22.39 & 10.66\tablenotemark{b} &   13.90 &      66 &      43 &          1 \\
                 Arp 248 NED02 &          Arp 248 &  5167 &   54.60 &   -23.32 & 10.66\tablenotemark{b} &   13.27 &  0$-$45 &     109 &          1 \\
Arp 248 NED03\tablenotemark{a} &          Arp 248 &  5276 &   93.10 &   -20.82 &                \nodata & \nodata & \nodata & \nodata &          1 \\
                      NGC 3994 &          Arp 313 &  3086 & \nodata &   -23.56 &                  10.55 &    6.18 &      58 &      10 &    1,2,3,4 \\
                      NGC 3995 &          Arp 313 &  3254 &   24.70 &   -22.64 &                  10.42 &   19.40 &      69 &      45 &    1,2,3,4 \\
     NGC 3991\tablenotemark{a} &          Arp 313 &  3192 &   50.10 &   -22.51 &                  10.26 & \nodata & \nodata & \nodata &          4 \\
                      NGC 7469 &          Arp 298 &  4892 & \nodata &   -24.23 &                  11.46 &   12.22 &      30 &     127 &      1,3,4 \\
                       IC 5283 &          Arp 298 &  4804 &   23.30 &   -23.36 &                \nodata &    9.90 &      60 &     105 &      1,3,4 \\
                     UGC 12545 &          WBL 713 &  5754 & \nodata &   -21.41 & 10.28\tablenotemark{b} &   11.81 &      63 &      85 &          2 \\
                     UGC 12546 &          WBL 713 &  6055 &   21.90 &   -22.69 & 10.28\tablenotemark{b} &   10.64 &      63 &      20 &          2 \\
  \enddata

  \tablerefs{1 = \citet{arp66a}; 2 = \citet{bgk00a}; 3 =
    \citet{sanders03a}; 4 = \citet{ssm04a}}

  \tablecomments{Col.(1): Galaxy name.  Col.(2): System label from
    optical identification.  In order of preference: Arp =
    \citet{arp66a}; VV = \citet{vv77a}; HCG = Hickson Compact Group =
    \citet{hickson82a}; WBL = \citet{white99a} catalog of poor
    clusters.  Col.(3): Redshift from NED, in \kms.  Col.(4):
    Projected nuclear separation from first galaxy listed for system,
    in kpc, using.  Col.(5): $K$-band magnitude down to 20
    mag/arcsec$^2$ isophote, from 2MASS.  Col.(6): 8$-$1000\micron\
    luminosity, determined using the formula in \citet{sm96a} and
    fluxes from \citet{ssm04a}, the {\it IRAS} Faint Source Catalog,
    or \citet{johnson07a} in the case of HCG 2.  Col.(7): $B$-band
    optical radius from 25 mag/arcsec$^2$ isophote, in kpc, from
    HyperLeda \citep{paturel03a}.  Col.(8): Inclination in degrees,
    from HyperLeda except for Arp 248 NED02 and Arp 256 NED02.  For
    these cases, we assume 30$^\circ$; the actual inclination is in
    the range $0-45^\circ$.  Col.(9): Position angle of the galaxy
    line of nodes, in degrees east of north; from HyperLeda except for
    Arp 256 and WBL 036 (used \hone\ data from \citealt{chen02a} and
    \citealt{condon02a}).  Col.(10): Sample selection reference
    (\S\ref{sec:sample}).}

  \tablenotetext{a}{No LRIS data is available for these galaxies.
    They are listed here as prominent but more distant members of the
    relevant systems (except for Mrk 552, which is the closest galaxy
    to UGC 312 in projection).}

  \tablenotetext{b}{The infrared luminosity for this
    pair is unresolved; the total system luminosity is computed using
    the redshift of each galaxy.}
  
\end{deluxetable}

\begin{deluxetable}{llrccrrccrrcc}
  \tabletypesize{\footnotesize}
  \tablecaption{Control Sample\label{tab:control}}
  \tablewidth{0pt}

  \tablehead{
    \colhead{Galaxy} & \colhead{Type} & \colhead{Distance} &
    \colhead{Dist. Ref.} & \colhead{$M_K$} & \colhead{log(\lir/\lsun)}
    & \colhead{\rtf} & \colhead{R$_d$} & \colhead{R$_d$ Ref.} &
    \colhead{$i$} & \colhead{PA$_{nodes}$} & \colhead{$i$/PA Ref.} &
    \colhead{\ion{H}{2} Ref.} \\
    \colhead{(1)} & \colhead{(2)} & \colhead{(3)} & \colhead{(4)} &
    \colhead{(5)} & \colhead{(6)} & \colhead{(7)} & \colhead{(8)} &
    \colhead{(9)} & \colhead{(10)} & \colhead{(11)} & \colhead{(12)} &
    \colhead{(13)}
  }

  \startdata
  NGC~~300        & SAd   &  2.08 & a & -19.43 &  8.43 &  5.90 & 2.26 & B & 40 & 114 & aa & 1 \\
  NGC~~628 (M74)  & SAc   &  8.59 & b & -22.48 &  9.82 & 13.08 & 3.57 & A &  7 &  25 & cc & 6,7 \\
  NGC~~925        & SABd  &  7.38 & e & -20.75 &  9.21 & 12.04 & 2.60 & D & 61 & 107 & aa & 7 \\
  NGC~1232        & SABc  & 14.50 & d & -23.08 &  9.91 & 14.59 & 4.51 & C & 33 & 270 & aa,ee & 3,7 \\
  NGC~2403        & SABcd &  3.16 & a & -21.05 &  9.19 &  9.17 & 1.72 & C & 60 & 127 & aa & 4,7 \\
  NGC~2805        & SABd  & 28.00 & c & -21.44 &  9.68 & 14.42 & 8.96 & E & 36 & 290 & aa,bb & 7 \\
  NGC~2903        & SABbc &  8.55 & e & -23.56 & 10.22 & 14.95 & 3.09 & E & 56 &  23 & aa & 3,7 \\
  NGC~2997        & SABc  &  7.08 & e & -22.61 &  9.82 & 11.55 & 4.84 & F & 41 & 110 & aa & 3 \\
  NGC~3184        & SABcd &  8.70 & c & -22.09 &  9.48 &  9.60 & 2.59 & G & 24 & 135 & aa & 7 \\
  NGC~5236 (M83)  & SABc  &  4.92 & a & -23.66 & 10.37 & 10.34 & 2.89 & F & 27 &  45 & aa & 2,3 \\
  NGC~5457 (M101) & SABcd &  6.96 & a & -23.26 & 10.23 & 29.19 & 5.19 & E & 37 &  18 & dd & 5 \\
  \enddata
  
  \tablerefs{{\bf Distances}: a = \citet{jacobs09a}; b =
    \citet{herrmann08a}; c = \citet{tully88a}; d = \citet{tully09a}; e
    = \citet{tully08a}.  {\bf \rtf}: A = \citet{boroson81a}; B =
    \citet{carignan85a}; C = \citet{elmegreen84a}; D =
    \citet{elmegreen85a}; E = \citet{koopmann06a}; F =
    \citet{simien86a}; G = \citet{vanzee98a}. {\bf $i$/PA}: aa =
    HyperLeda (\citealt{paturel03a}); bb = \citet{bosma80a}; cc =
    \citet{kamphuis92a}; dd = \citet{kennicutt96a}; ee =
    \citet{vanzee99a}. {\bf \ion{H}{2} regions}: 1 =
    \citet{bresolin09b}; 2 = \citet{bresolin09a}; 3 =
    \citet{bresolin05a}; 4 = \citet{garnett97a}; 5 = \citet{kbg03a}; 6
    = \citet{mccall85a}; 7 = \citet{vanzee98a}.}

  \tablecomments{Col.(1): Galaxy name.  Col.(2): Morphological type,
    from NED.  Col.(3-4): Distance, in Mpc, and reference.  Col.(5):
    $K$-band magnitude down to 20 mag/arcsec$^2$ isophote, from 2MASS.
    Col.(6): 8$-$1000\micron\ luminosity, determined using the formula
    in \citet{sm96a} and infrared fluxes from NED.  Col.(7): $B$-band
    optical radius from 25 mag/arcsec$^2$ isophote, in kpc, from
    HyperLeda (\citealt{paturel03a}; except for M101, which is from
    the RC3).  Col.(8-9): $B$-band stellar exponential disk scale
    length, in kpc, and reference.  Col. (10-12): Inclination, in
    degrees; position angle of the galaxy line of nodes, in degrees
    east of north; and reference for these.  Col.(13): Reference for
    emission-line data.}

\end{deluxetable}

\begin{deluxetable}{lrlll}
  \tablecolumns{5}
  \tablecaption{Gradient Fits \label{tab:gradients}}
  \tablewidth{0pt}

  \tablehead{ \colhead{Galaxy} & \colhead{$N_{reg}$} &
    \colhead{$\Delta$(dex/kpc)} & \colhead{$\Delta$(dex/\rtf)} &
    \colhead{Intercept} \\
    \colhead{(1)} & \colhead{(2)} & \colhead{(3)} & \colhead{(4)} &
    \colhead{(5)}}

  \startdata
\cutinhead{Control Sample}
NGC  300        &  27 &             -0.1246 &              -0.664 &              8.95 \\
NGC  628        &  25 &             -0.0410 &              -0.536 &              9.10 \\
NGC  925        &  44 &             -0.0471 &              -0.568 &              8.83 \\
NGC 1232        &  22 &             -0.0310 &              -0.451 &              9.10 \\
NGC 2403        &  25 &             -0.0750 &              -0.689 &              8.93 \\
NGC 2805        &  15 &             -0.0237 &              -0.342 &              8.93 \\
NGC 2903        &  17 &             -0.0239 &              -0.357 &              9.08 \\
NGC 2997        &  14 &             -0.0642 &              -0.741 &              9.30 \\
NGC 3184        &  17 &             -0.0613 &              -0.590 &              9.35 \\
NGC 5236        &  38 &             -0.0413 &              -0.427 &              9.29 \\
NGC 5457        &  25 &             -0.0310 &              -0.899 &              9.13 \\
\cutinhead{Interacting Galaxies Sample}
UGC 12914       &  20 & -0.0071$\pm$ 0.0033 &  -0.104$\pm$  0.048 &   9.13$\pm$  0.02 \\
Arp 256 NED01   &  13 & -0.0142$\pm$ 0.0022 &  -0.193$\pm$  0.031 &   8.95$\pm$  0.01 \\
Arp 256 NED02   &  25 & -0.0162$\pm$ 0.0014 &  -0.271$\pm$  0.023 &   9.05$\pm$  0.01 \\
UGC 312         &  13 & -0.0114$\pm$ 0.0070 &  -0.133$\pm$  0.081 &   8.84$\pm$  0.05 \\
UGC 813         &  20 & -0.0210$\pm$ 0.0040 &  -0.233$\pm$  0.045 &   9.00$\pm$  0.02 \\
UGC 816         &  21 & -0.0189$\pm$ 0.0023 &  -0.252$\pm$  0.030 &   9.05$\pm$  0.02 \\
NGC 2207        &  43 & -0.0124$\pm$ 0.0026 &  -0.273$\pm$  0.057 &   9.08$\pm$  0.03 \\
IC 2163         &  19 & -0.0165$\pm$ 0.0036 &  -0.162$\pm$  0.035 &   9.17$\pm$  0.02 \\
Arp 248 NED01   &   8 & -0.0368$\pm$ 0.0040 &  -0.512$\pm$  0.056 &   8.85$\pm$  0.02 \\
Arp 248 NED02   &  11 & -0.0144$\pm$ 0.0020 &  -0.190$\pm$  0.026 &   9.03$\pm$  0.01 \\
NGC 3994        &  25 & -0.0224$\pm$ 0.0072 &  -0.139$\pm$  0.045 &   9.08$\pm$  0.02 \\
NGC 3995        &  41 & -0.0175$\pm$ 0.0030 &  -0.339$\pm$  0.059 &   8.86$\pm$  0.03 \\
NGC 7469        &  12 & -0.0146$\pm$ 0.0089 &  -0.179$\pm$  0.109 &   9.09$\pm$  0.05 \\
IC 5283         &  12 & -0.0249$\pm$ 0.0060 &  -0.246$\pm$  0.059 &   9.14$\pm$  0.01 \\
UGC 12545       &  14 & -0.0323$\pm$ 0.0142 &  -0.381$\pm$  0.168 &   8.87$\pm$  0.08 \\
UGC 12546       &  20 & -0.0395$\pm$ 0.0082 &  -0.421$\pm$  0.087 &   9.10$\pm$  0.03 \\
  \enddata

  \tablecomments{Col.(1): Galaxy name.  Col.(2): Number of \ion{H}{2}
    regions used in fit.  Col.(3-4): Radial oxygen abundance gradient,
    in dex/kpc and dex/\rtf.  Col.(5).  Intercept of oxygen abundance
    vs. radius fit, expressed as 12$+$log(O/H).}

\end{deluxetable}

\end{document}